\documentclass[aps,prd,12pt,notitlepage,showpacs,nofootinbib,tightenlines]{revtex4-1}
\usepackage{amsmath}
\usepackage{amssymb}
\usepackage{epsfig}
\usepackage{graphicx}
\usepackage{bm}
\usepackage{times}
\usepackage{braket}
\usepackage{color}
\usepackage{slashed}
\usepackage{hyperref}
\definecolor{Red}{rgb}{1.,0.,0.}

\definecolor{Blue}{rgb}{0.,0.,1.}

\definecolor{nicered}{rgb}{0.7,0.1,0.1}
\definecolor{nicegreen}{rgb}{0.1,0.5,0.1}
\hypersetup{colorlinks,citecolor=nicegreen,linkcolor=nicered}

\begin{document}
%%%%%%%%%%%%%%%%%%%%%%%%%%%%%%%%%%%%%%%%%%%%%

\newcommand{\beq}{\begin{eqnarray}}
\newcommand{\eeq}{\end{eqnarray}}
\newcommand{\non}{\nonumber\\ }

\newcommand{\jpsi}{J/\Psi}

\newcommand{\ppa}{\phi_\pi^{\rm A}}
\newcommand{\ppp}{\phi_\pi^{\rm P}}
\newcommand{\ppt}{\phi_\pi^{\rm T}}
\newcommand{\ov}{ \overline }

\newcommand{\zerot}{ {\textbf 0_{\rm T}} }
\newcommand{\kt}{k_{\rm T} }
\newcommand{\fb}{f_{\rm B} }
\newcommand{\fk}{f_{\rm K} }
\newcommand{\rk}{r_{\rm K} }
\newcommand{\mb}{m_{\rm B} }
\newcommand{\mw}{m_{\rm W} }
\newcommand{\im}{{\rm Im} }

\newcommand{\kks}{K^{(*)}}
\newcommand{\acp}{{\cal A}_{\rm CP}}
\newcommand{\pb}{\phi_{\rm B}}

\newcommand{\xeba}{\bar{x}_2}
\newcommand{\xsba}{\bar{x}_3}
\newcommand{\peas}{\phi^A}

\newcommand{\pvsl}{ p \hspace{-2.0truemm}/_{K^*} }
\newcommand{\esl}{ \epsilon \hspace{-2.1truemm}/ }
\newcommand{\psl}{ p \hspace{-2truemm}/ }
\newcommand{\ksl}{ k \hspace{-2.2truemm}/ }
\newcommand{\lsl}{ l \hspace{-2.2truemm}/ }
\newcommand{\nsl}{ n \hspace{-2.2truemm}/ }
\newcommand{\vsl}{ v \hspace{-2.2truemm}/ }
\newcommand{\epsl}{\epsilon \hspace{-1.8truemm}/\,  }
\newcommand{\bfkk}{{\bf k} }
\newcommand{\calm}{ {\cal M} }
\newcommand{\calh}{ {\cal H} }

%%---------------------------------------------------------
%%%%%%%%%%%%%%%%%%%
\def \appb{{\bf Acta. Phys. Polon. B }  }
\def \cpc{ {\bf Chin. Phys. C } }
\def \ctp{ {\bf Commun. Theor. Phys. } }
\def \epjc{{\bf Eur. Phys. J. C} }
\def \jhep{{\bf J. High Energy Phys. } }
\def \jpg{ {\bf J. Phys. G} }
\def \mpla{{\bf Mod. Phys. Lett. A } }
\def \npb{ {\bf Nucl. Phys. B} }
\def \plb{ {\bf Phys. Lett. B} }
\def \pr{  {\bf Phys. Rep.} }
\def \prc{ {\bf Phys. Rev. C }}
\def \prd{ {\bf Phys. Rev. D} }
\def \prl{ {\bf Phys. Rev. Lett.}  }
\def \ptp{ {\bf Prog. Theor. Phys. }}
\def \zpc{ {\bf Z. Phys. C}  }

%%%%%%%%%%%%%%%%%%%%%%%%%%%%%%%%%%%%%%%%%%%%%%%%%%%%
%%
\title{The NLO twist-3 contribution to the pion electromagnetic form
factors in $k_{T}$ factorization}
\author{Shan Cheng} %\email{chengshan-anhui@163.com}
\author{Ying-Ying Fan} %\email{fyy163@126.com}
\author{Zhen-Jun Xiao }\email{xiaozhenjun@njnu.edu.cn}
\affiliation{  Department of Physics and Institute of Theoretical Physics,
Nanjing Normal University, Nanjing, Jiangsu 210023, People's Republic of China,}
\date{\today}
\vspace{1cm}
\begin{abstract}
In this paper, by employing  the $k_{T}$ factorization theorem, we calculate firstly
the next-to-leading-order (NLO) twist-3 contributions to the pion electromagnetic
form factors in the $\pi\gamma^* \to \pi$ process.
From the analytical and numerical calculations we find the following points:
(a) For the leading order (LO) twist-2, twist-3 and the NLO twist-2 contributions,
our results agree very well
with those obtained in previous works;
(b) We extract out two factors $F^{(1)}_{\rm T3}(x_i,t,Q^2)$ and
$\ov{F}^{(1)}_{\rm T3}(x_i,t,Q^2)$,
which describe directly the NLO twist-3 contributions to the pion
electromagnetic form factors $F^+(Q^2)$;
(c) The NLO twist-3 contribution is negative in sign and cancel
partially with the NLO twist-2
part, the total NLO contribution can therefore provide a roughly
$\pm 20\%$ corrections to the total LO
contribution in the considered ranges of $Q^2$;
and (d) The theoretical predictions for $Q^2 F^+(Q^2)$ in the
low-$Q^2$ region agree well with currently available
data, this agreement can be improved by the inclusion of the NLO contributions.
\end{abstract}

%\vspace{1cm}

%%\keywords{$\kt$ factorization, next-to-leading-order correction, $\pi$ form factor, twist-3.}

\maketitle

\section{Introduction}

The perturbative QCD (pQCD) factorization approach, based on the $\kt$ factorization
theorem\cite{npb325-62,npb360-3,prl74-4388},
have been wildly used to deal with the inclusive and exclusive processes
\cite{plb504-6,prd63-074009,plb242-97,prl65-2343}.
In the  $\kt$ factorization theorem, the end-point singularities
are removed by the small but non-zero transverse momentum $\kt$ of the parton
propagators. For many years, the application of the $k_{T}$ factorization theorem
were mainly at the leading order (LO) level. But the  situation changed a lot recently.
In Refs.~\cite{prd76-034008,prd83-054029,prd85-074004}, the authors
calculated the next-to-leading order (NLO) twist-2 contributions
to the $\pi$ transition form factor, $\pi$ electromagnetic
form factor and $B \to \pi$ form factor respectively, obtained
the infrared finite $k_{T}$ dependent NLO hard kernel, and therefore
confirmed the applicability of the $k_{T}$ factorization to these
exclusive processes at the NLO and the leading twist (twist-2) level.
This fact tell us that the $k_{T}$ factorization approach can also be applied
to the high order contributions as mentioned in Ref.~\cite{prd64-014019}.

In the framework of the pQCD factorization approach,
the contributions to the form factors include four parts:
\begin{enumerate}
\item[(i)]
The leading order contribution include the leading order twist-2 (LO-T2)
contribution and the leading order twist-3 (LO-T3) contribution.

\item[(ii)]
The NLO contribution contains the NLO twist-2 (NLO-T2) contribution and
the NLO  twist-3 (NLO-T3) contribution.

\end{enumerate}
At present, the first three parts, namely the LO-T2, LO-T3 and NLO-T2 contributions,
have been evaluated in Refs.~\cite{prd76-034008,prd83-054029,prd85-074004},
but the NLO-T3 contribution is still absent now.

At leading order level,  the LO-T2 part is smaller than the LO-T3 part,
by a ratio of $\sim 34\%$ against $\sim 66\%$ as shown in
Refs.~\cite{prd65-014007,epjc23-275,prd83-054029}.
The NLO-T2 part is around $20-30\%$ of the total leading order contribution (i.e.
LO-T2 plus LO-T3 part ) in the low $Q^2$ region.
Since the LO-T3 contribution is large, the remaining unknown
fourth part, the NLO twist-3 contribution, maybe rather important, and
should be calculated in order to obtain the pQCD predictions for relevant form factors
at the full NLO level, and to demonstrate that the $k_{T}$
factorization theorem is an systematical tool.

In this paper we concentrate on the calculation for the NLO twist-3 contribution
to the $\pi$ electromagnetic form factor, which corresponds to the scattering process
$\pi \gamma^{\star}\to \pi$. Our work represents the first calculation for
the NLO  twist-3 contribution to this quantity in the $k_{T}$ factorization theorem.

We know that the collinear divergences would appear when the massless gluon is
emitted from the light external line as the gluon is paralleled to the initial-
or the final-state pion which are massless assumed.
The soft divergences would come from the exchange of the massless gluon between
two on-shell external lines. In this work light partons are considered to be
off-shell by $k_{T}^{2}$ to regulated the infrared divergences
in both the QCD quark diagrams and the effective diagrams for pion wave functions.
It's a nontrivial work to verify that the collinear divergences
from the quark-level diagrams offset those from the pion wave functions and
the soft divergences cancel among quark-level diagrams exactly
at the twist-3 level as well as at the leading twist-2 case \cite{prd83-054029}.
As demonstrated in Refs.~\cite{prd83-054029,prd85-074004}, both the
large double logarithms $\alpha_{s}\ln^{2}(k_{T})$ and  $\alpha_{s}\ln^{2}(x_{i})$,
here $x_{i}$ being the parton momentum fraction of the anti-quark
in the meson wave functions,
could be absorbed through the resummation technology.
The double logarithm $\alpha_{s}\ln^{2}(k_{T})$ would be absorbed
into the $\pi$ meson wave functions and then
been summed to all orders in the coupling constant $\alpha_{s}$ by the $k_{T}$
resummation\cite{prl74-4388}.
The jet function would included when there exist the end-point
singularity in the hard kernel,
and then the double logarithm $\alpha_{s}\ln^{2}(x_{i})$ would
be summed to all orders by the threshold resummation\cite{prd66-094010,plb555-197}.
The renormalization scale $\mu$ and the factorization scale
$\mu_{f}$ are introduced in the high-order corrections to the QCD quark diagrams
and the effective diagrams,  respectively.
With the appropriate choice of the scale $\mu$ and $\mu_{f}$,
say setting them as the internal hard scale as postulated in \cite{prd83-054029},
the NLO correction are under control.

This paper is organized as follows. In section.~II, we give a brief review
about the evaluations of the LO diagrams for the process $\pi \gamma^*\to \pi$,
for both the
twist-2 part and twist-3 part. In section.~III, $O(\alpha^{2}_{s})$ QCD
quark diagrams for the
process will be calculated with the inclusion of the twist-3 contributions.
The convolutions of $O(\alpha_{s})$(~NLO) effective diagrams for
the meson wave functions and $O(\alpha_{s})$(~LO) hard kernel would also
be presented in this section,
then the $k_{T}$-dependent NLO hard kernel at twist-3 will be obtained.
Section.~IV contains the numerical analysis.
With appropriate choices for the renormalization scale $\mu$, the
factorization scale $\mu_f$ and the input meson wave functions,
we make the numerical calculations for all four parts of the LO and NLO
contributions to the pion
electromagnetic form factor in the $\pi \gamma^* \to \pi$ process.
Section V contains the conclusions.

%%%%%%%%%%%%%%%%%%%%%%%%%%%%%%%%%%%%%%%%%%%%%%%

\section{LO twist-2 and twist-3 contributions}

\begin{figure}[tb]
\vspace{-1cm}
\begin{center}
\leftline{\epsfxsize=9cm\epsffile{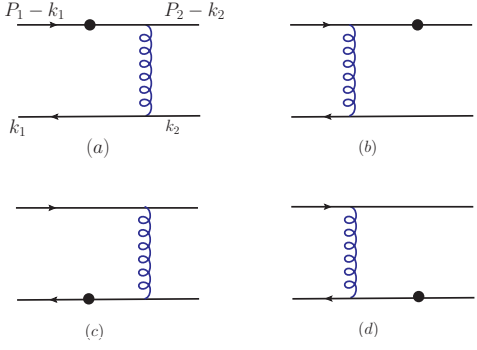}}
\end{center}
\vspace{-8cm}
\caption{Leading-order quark diagrams for the $\pi \gamma^{\star} \to \pi$
form factor with $\bullet$ representing the virtual photon vertex.}
\label{fig:fig1}
\end{figure}

The leading order hard kernels of the $\pi$ electromagnetic form
factor as shown in Fig.~\ref{fig:fig1} are calculated in this section.
The $\pi \gamma^{\star} \to \pi$ form factors are defined via the matrix element
\beq
<\pi(p_{2})| J^{\mu} |\pi(p_{1})>  &=&  f_{1}(q^2)p^{\mu}_{1} + f_{2}(q^2)p^{\mu}_{2}\non
 &=& F^{+}(q^{2})(p^{\mu}_{1}+p^{\mu}_{2}), \label{eq:ffme}
\eeq
where $p_1$ ($p_2$) refers to the momentum of the initial (final) state
pion, $q = p_1 - p_2$ is the momentum transferred in the weak vertex.
Using the same definitions for the leading case as
being used in Ref.~\cite{prd83-054029}, the momentum $p_1$ and $p_2$ are  chosen as
\beq
p_{1} = (p^{+}_{1},0,\zerot ), \quad
p_{2} = (0,p^{-}_{2},\zerot),
\eeq
with $q^{2} = -2 p_{1} \cdot p_{2} = - Q^{2}$.
According to the $\kt$ factorization, the $k_{1} = (x_{1}p^{+}_{1}, 0 , k_{\rm 1T})$
in the initial pion meson and $k_{2} = (0, x_{2}p^{-}_{2}, k_{\rm 2T})$
in the final pion meson as labeled in Fig.~\ref{fig:fig1}, and
$x_{1}$ and $x_{2}$ being the momentum fractions.
The follow hierarchy is postulated in the small-x region:
\beq
Q^2 \gg x_2 Q^2 \sim x_1 Q^2 \gg x_1 x_2 Q^2 \gg k^2_{1T}, k^2_{2T},
\label{eq:hierarchy}
\eeq

The following Fierz identity is employed to factorize the fermion flow.
\beq
I_{ij}I_{lk} &=& \frac{1}{4}I_{ik}I_{lj} + \frac{1}{4}(\gamma_{5})_{ik}(\gamma_{5})_{lj}
               + \frac{1}{4}(\gamma^{\alpha})_{ik}(\gamma^{\alpha})_{lj}\non
             &&+ \frac{1}{4}(\gamma_{5}\gamma^{\alpha})_{ik}(\gamma_{\alpha}\gamma_{5})_{lj}
               + \frac{1}{8}(\sigma^{\alpha\beta}\gamma_{5})_{ik}
               (\sigma_{\alpha\beta}\gamma_{5})_{lj}.
\label{eq:fierz}
\eeq
The identity matrix I here is a $4$ dimension matrix,
the structure $\gamma_{\alpha}\gamma_{5}$ in Eq.~(\ref{eq:fierz}) contribute at the
leading twist(twist-2),
while $\gamma_{5}$ and $\sigma_{\alpha\beta} \gamma_{5}$ contribute at twist-3 level.
The identity of $SU(3)_c$ group,
\beq
I_{ij}I_{lk} = \frac{1}{N_{c}}I_{lj}I_{ik} + 2(T^{c})_{lj}(T^{c})_{ik}
\label{eq:color}
\eeq
is also employed to factorize the color flow.
In Eq.~(\ref{eq:color}), $(i,j,l,k)$ are color index, $N_{c}=3$ is the
number of the colors, and $T^{c}$ is the Gel-Mann color matrix of $SU(3)_c$.
The first term in Eq.~(\ref{eq:color}) corresponds to the color-singlet
state of the valence quark
and the anti-quark, while the second term will be associated with the color-octet state.

We here consider only the subdiagram Fig.~\ref{fig:fig1}(a) in detail,
where the quark and anti-quark form a color-singlet state.
The hard kernels of the other subdiagrams can be obtained by simply kinetic replacements.
The wave function $\Phi_\pi(p_i,x_i)$ for the initial and final state
pion can be written as the following form \cite{zpc48-239,jhep9901-010,jhep0605-004}
\beq
\Phi_\pi(p_1,x_1)&=& \frac{i}{\sqrt{2N_{c}}} \gamma_5 \left \{ \psl_1 \phi_\pi^A(x_1)
+ m_0 \left [ \phi^{P}_{\pi}(x_1) - (\nsl_{+}\nsl_{-} - 1 )\phi^{T}_{\pi}(x_{1}) \right ]
\right\},  \label{eq:phi01}\\
\Phi_\pi(p_2,x_2)&=& \frac{i}{\sqrt{2N_{c}}} \gamma_5 \left \{ \psl_2 \phi_\pi^A(x_2)
+ m_0 \left [ \phi^{P}_{\pi}(x_2) - ( \nsl_{-}\nsl_{+} - 1 )\phi^{T}_{\pi}(x_2) \right ]
\right\},  \label{eq:phi02}
\eeq
where $n_{+} = (1,0,\zerot)$ and $n_{-} = (0,1,\zerot)$ denote the unit vector
along with the positive and negative $z$-axis direction, $m_0=1.74$ GeV is the chiral mass
of pion,  $N_{c}$ is the number of colors, $\phi_\pi^A(x_i)$ are the leading
twist-2 pion distribution amplitude,
while $\phi_\pi^P(x_i)$ and $\phi_\pi^T(x_i)$ are the twist-3 pion distribution
amplitudes.

Combining the decompositions in  Eq.~(\ref{eq:fierz}) and Eq.~(\ref{eq:color}),
we then can sandwich Fig.~\ref{fig:fig1}(a) with the structures
\beq
\frac{1}{4N_{c}}\psl_{1}\gamma_{5}, \quad\frac{1}{4N_{c}}\gamma_{5}\psl_{2},
\label{eq:sanwich-tw2}
\eeq
from the initial and final state respectively,
in order to obtain the hard kernel $H^{(0)}$ at twist-2 level.
For the derivation of the twist-3 hard kernel, one should sandwich
Fig.~\ref{fig:fig1}(a) with the following two sets of structures
\beq
\left (\frac{1}{4N_{c}}\gamma_{5},\quad \frac{1}{4N_{c}}\gamma_{5}\right); \qquad
\left ( \frac{1}{8N_{c}}\sigma^{\alpha\beta}\gamma_{5},
\quad \frac{1}{8N_{c}}\sigma_{\alpha\beta}\gamma_{5}\right).
\label{eq:sanwich-tw3}
\eeq

Then the LO twist-3 contribution to the hard kernel from Fig.~\ref{fig:fig1}(a)
can be written as\cite{prd84-034018}
\beq
H^{(0)}_{a}(x_{1},k_{\rm 1T},x_{2},k_{\rm 2T})&=&(-2ie_q) 4\pi \alpha_s
\frac{C_F}{16N_{c}} m^{2}_{0} \phi^{P}_{\pi}(x_{2})\non
&& \hspace{-3cm}\cdot \Bigl \{ \frac{- 4 p^{\mu}_{2}
\left [\phi^{P}_{\pi}(x_{1}) - \phi^{T}_{\pi}(x_{1}) \right ]}
     {(p_{2}-k_{1})^{2}(k_{1}-k_{2})^{2}}
+ \frac{ 4x_{1} p^{\mu}_{1} \left [\phi^{P}_{\pi}(x_{1}) + \phi^{T}_{\pi}(x_{1}) \right ]}
     {(p_{2}-k_{1})^{2}(k_{1}-k_{2})^{2}} \Bigr\},
\label{eq:lot3hka}
\eeq
where $\alpha_s$ in the strong coupling constant, $C_F=4/3$ in s color factor,
$e_q$ refers to the charge of the quark interacting with the $\gamma^*$ in
$\pi \gamma^* \to \pi$ process.

The corresponding LO twist-2 contribution to the hard kernel takes the form of
\beq
H^{(0)}_{\rm a,T2}(x_{1},k_{1T},x_{2},k_{2T})=(ie_q)
4\pi \alpha_s\; \frac{C_F}{16N_{c}} Q^{2} \phi^{A}_{\pi}(x_{2}) \phi^{A}_{\pi}(x_{1})
\cdot  \frac{4 x_1 p^{\mu}_{1}}{(p_{1}-k_{2})^{2}(k_{1}-k_{2})^{2}},
\label{eq:lot2hka}
\eeq

%%-----------------------------------------------------------------------
%%-----------------------------------------------------------------------

It is easy to see that all parts of the initial state pion, the twist-2 $\phi_\pi^A(x_1)$
and twist-3 $\phi_\pi^P(x_1)$ and $\phi_\pi^T(x_1)$, provide contributions at
leading order level,
but only the $\phi_\pi^A(x_2)$ and $\phi_\pi^P(x_2)$ of the final state pion
contribute at LO level, because the contribution from the $\phi_\pi^T(x_2)$ component
become zero when it is contracting  with the gluon propagator.
For the LO twist-3 hard kernel $H^{(0)}_{a}(x_{1},k_{1T},x_{2},k_{2T})$,
one can see that it contains two lorentz structures: $p^{\mu}_{2}$ term
and $x_{1}p^{\mu}_{1}$ term, these two terms all should be included in the
numerical calculations.
For the LO twist-2 hard kernel $H^{(0)}_{\rm a, T2}$ as given in Eq.~(\ref{eq:lot2hka}),
it depends on one term $x_{1}p^{\mu}_{1}$ only.
From previous studies in Ref.\cite{prd65-014007,epjc23-275,prd83-054029},
we know that the LO twist-2 part is only
about half of the LO twist-3 part. So one generally
expect that the NLO twist-3 contribution maybe large and essential
for considered transitions, which is also one of the motivations of this paper.

\section{NLO corrections}

Under the hierarchy as  shown in Eq.~(\ref{eq:hierarchy}),  only those terms which
don't vanish in the limits of $x \to 0$ and $k_{iT} \to 0$ should be  kept, this
fact does simplify the expressions of the NLO contributions greatly.

From the discussions at the end of Sec.~I, we know that both lorentz structures
$x_1p_1^\mu$ and $p_2^\mu$ will contribute.
From the hard kernel  $H_a^{(0)}(x_{1},k_{\rm 1T},x_{2},k_{\rm 2T} )$ as given in
Eq.~(\ref{eq:lot3hka}), we define those two parts of the LO twist-3
contribution, $H_a^{(0)}(x_{1}p^{\mu}_{1})$ and $H_a^{(0)}(p^{\mu}_2)$ in the form of
\beq
H_a^{(0)}(x_{1}p^{\mu}_{1})  &\equiv& (-2ie_q) 4\pi \alpha_s
\frac{C_F}{16N_{c}} m^{2}_{0} \phi^{P}_{\pi}(x_{2}) \frac{4 x_{1}
p^{\mu}_{1} [\phi^{P}_{\pi}(x_{1}) - \phi^{T}_{\pi}(x_{1})]}{(p_{2}
-k_{1})^{2}(k_{1}-k_{2})^{2}},
\label{eq:lot3hka1}\\
H_a^{(0)}(p^{\mu}_{2})  &\equiv& (-2ie_q) 4\pi \alpha_s
\frac{C_F}{16N_{c}} m^{2}_{0} \phi^{P}_{\pi}(x_{2}) \frac{- 4 p^{\mu}_{2}
[\phi^{P}_{\pi}(x_{1}) + \phi^{T}_{\pi}(x_{1})]}{(p_{2}
-k_{1})^{2}(k_{1}-k_{2})^{2}},\label{eq:lot3hka2}\\
H_a^{(0)}  &=& H_a^{(0)}(x_{1}p^{\mu}_{1}) + H_a^{(0)}( p^{\mu}_2).\label{eq:ha0-t3}
\eeq
For Figs.~\ref{fig:fig1}(b,c,d), one can find the corresponding LO twist-3
contributions by simple replacements.  For the sake of simplicity, we
will generally omit the subscript ``$a$" in $H_a^{(0)}$ in the following sections,
unless stated specifically.

\subsection{NLO twist-3 Contributions of the QCD Quark Diagrams}

\begin{figure}[]
\vspace{-1cm}
\begin{center}
\leftline{\epsfxsize=10cm\epsffile{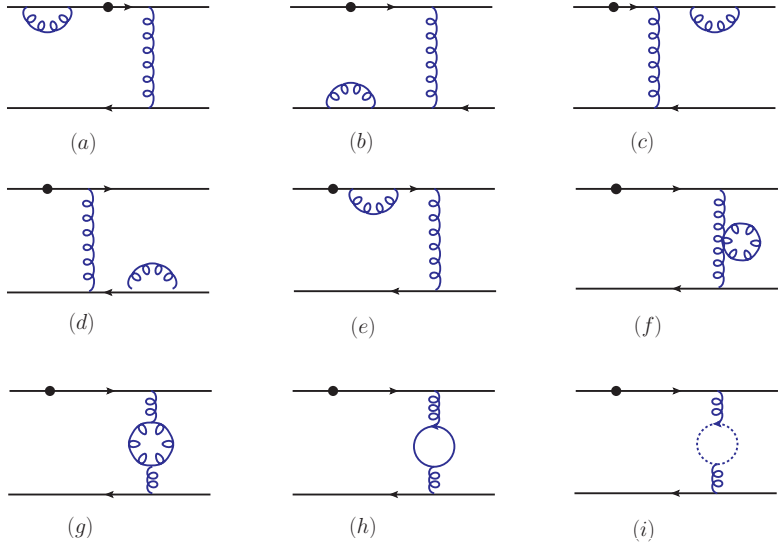}}
\end{center}
\vspace{-6cm}
\caption{Self-energy corrections to Fig.1(a).}
\label{fig:fig2}
\end{figure}

\begin{figure}[]
\vspace{-1cm}
\begin{center}
\leftline{\epsfxsize=10cm\epsffile{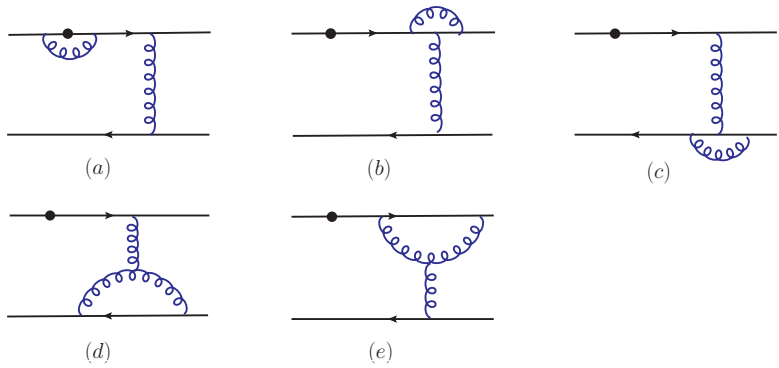}}
\end{center}
\vspace{-8cm}
\caption{Vertex corrections to Fig.1(a).}
\label{fig:fig3}
\end{figure}

\begin{figure}[]
\vspace{-1cm}
\begin{center}
\leftline{\epsfxsize=8cm\epsffile{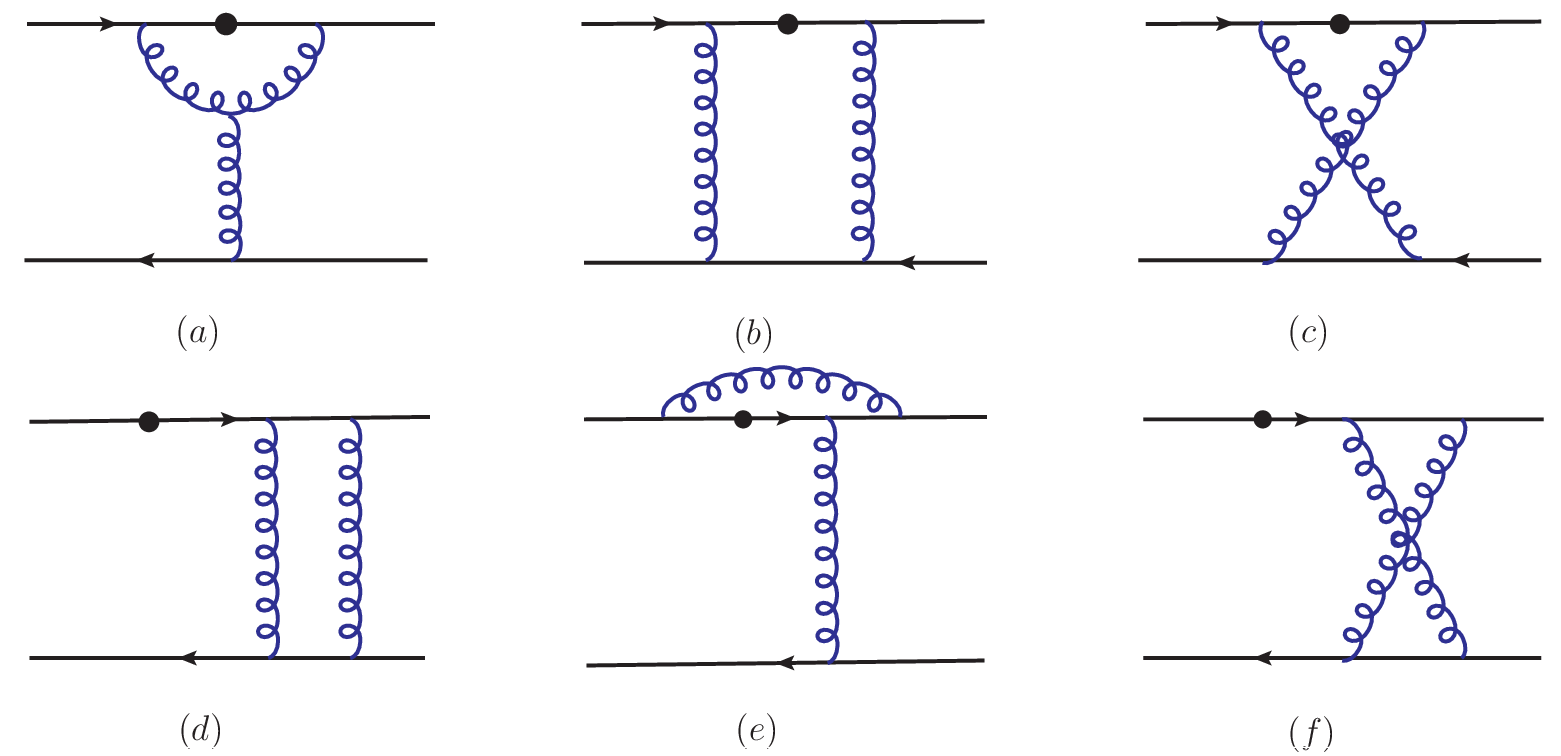}}
\end{center}
\vspace{-7cm}
\caption{Box and pentagon corrections to Fig.1(a).}
\label{fig:fig4}
\end{figure}

Now we calculate the NLO twist-3 contributions to Fig.~\ref{fig:fig1}(a), which
comes from the self-energy diagrams, the vertex diagrams, the box and the pentagon
diagrams, as illustrated in Figs.~\ref{fig:fig2},\ref{fig:fig3} and \ref{fig:fig4}
respectively. After completing the calculations for Fig.~\ref{fig:fig1}(a), we
can obtain the results for other three figures: Fig.~\ref{fig:fig1}(b,c,d), by
simple replacements.

The ultraviolet(UV) divergences are extracted in the dimensional reduction
\cite{plb84-193} in order to
avoid the ambiguity from handing the matrix $\gamma_{5}$.
The infrared(IR) divergences are identified as the logarithms $\ln{\delta_1}$ ,
$\ln{\delta_2}$ and their combinations, as defined in Ref.~\cite{prd83-054029}
\beq
\delta_1 = \frac{k^2_{1T}}{Q^2}, \quad
\delta_2 = \frac{k^2_{2T}}{Q^2}, \quad
\delta_{12} = \frac{-(k_1 - k_2)^2}{Q^2}. \label{eq:defi1}
\eeq

The self-energy corrections obtained by evaluating the one-loop
Feynman diagrams in Fig.~\ref{fig:fig2}(a-f) are of the form
\beq
&&G^{(1)}_{2a} = -\frac{\alpha_s C_F}{8 \pi}\left [\frac{1}{\epsilon} +
\ln{\frac{4 \pi \mu^2}{\delta_1 Q^2 e^{\gamma_E}}} + 2 \right ]H^{(0)},\non
&&G^{(1)}_{2b} = -\frac{\alpha_s C_F}{8 \pi}\left [\frac{1}{\epsilon} +
\ln{\frac{4 \pi \mu^2}{\delta_1 Q^2 e^{\gamma_E}}} + 2 \right ]H^{(0)},
\eeq
\beq
&&G^{(1)}_{2c} = -\frac{\alpha_s C_F}{8 \pi}\left [\frac{1}{\epsilon} +
\ln{\frac{4 \pi \mu^2}{\delta_2 Q^2 e^{\gamma_E}}} + 2 \right ]H^{(0)},\non
&&G^{(1)}_{2d} = -\frac{\alpha_s C_F}{8 \pi}\left [\frac{1}{\epsilon} +
\ln{\frac{4 \pi \mu^2}{\delta_2 Q^2 e^{\gamma_E}}} + 2 \right ]H^{(0)},\non
&&G^{(1)}_{2e} = -\frac{\alpha_s C_F}{4 \pi} \left [\frac{1}{\epsilon} +
\ln{\frac{4 \pi \mu^2}{x_1 Q^2 e^{\gamma_E}}} + 2 \right ]H^{(0)},\non
&&G^{(1)}_{2f+2g+2h+2i} = \frac{\alpha_s}{4 \pi} \left [
\left (\frac{5}{3} N_c -\frac{2}{3} N_f \right )
\left  (\frac{1}{\epsilon} + \ln{\frac{4 \pi \mu^2}{\delta_{12}
Q^2 e^{\gamma_E}}} \right ) \right ]H^{(0)},
\label{eq:self1}
\eeq
where  $1 / \epsilon$ represents the UV pole term, $\mu$ is the renormalization
scale, $\gamma_E$ is the Euler constant, $N_c$ is the number of quark color,
$N_f$ is the number of the active quarks flavors, and $H^{(0)}$ denotes the
LO twist-3 hard kernel described in Eq.~(\ref{eq:lot3hka}).
The Fig.~\ref{fig:fig2}(f,g,h,i) denotes the self-energy
correction to the exchanged gluon.

%%-----------------------------------------------------------------------
%%-----------------------------------------------------------------------

%-----------------------------------------------------------------------
%%-----------------------------------------------------------------------

It is easy to find that, the NLO self-energy corrections to the LO twist-3
hard kernels as listed in Eq.~(\ref{eq:self1}) are identical
in form to those self-energy corrections to the LO twist-2
hard kernels as given in Eqs.~(6-9) in Ref.~\cite{prd83-054029}.
The reason is that the self-energy diagrams don't involve the external lines
and therefore are irrelevant with the twist structures of the wave functions.
It should be note that an additional symmetry factor $\frac{1}{2}$  appeared
from the choice of the gluon endpoint to attach the external line in the
self-energy correction Fig.~\ref{fig:fig2}(a,b,c,d).
The self-energy corrections to the external lines will be canceled by the
responding effective diagrams as shown in Fig.~(\ref{fig:fig5},\ref{fig:fig6}).
The self-energy correction to the internal quark line as shown in
Fig.~\ref{fig:fig2}(e) don't generate any IR divergences.

The vertex corrections obtained by evaluating the one-loop
Feynman diagrams in Fig.~\ref{fig:fig3}(a-e) are of the form
\beq
G^{(1)}_{3a} &=&  \frac{\alpha_s C_F}{4 \pi}
\left [\frac{1}{\epsilon} + \ln{\frac{4 \pi \mu^2}{Q^2 e^{\gamma_E}}} +
\frac{1}{2} \right ]H^{(0)}, \non
G^{(1)}_{3b} &=& -\frac{\alpha_s }{8 \pi N_c} \left [\frac{1}{\epsilon} +
\ln{\frac{4 \pi \mu^2}{x_1 Q^2 e^{\gamma_E}}} - 1 \right ]H^{(0)}\non
&&      -\frac{\alpha_s }{8 \pi N_c} \left [1 - \ln{\frac{\delta_{2}}{x_{1}}}
\right ]H^{(0)}(x_{1}p^{\mu}_{1}),\non
G^{(1)}_{3c} &=& -\frac{\alpha_s}{8 \pi N_c}
\left [\frac{1}{\epsilon} + \ln{\frac{4 \pi \mu^2}{\delta_{12} Q^2 e^{\gamma_E}}}
\right ]H^{(0)} \non
&& -  \frac{\alpha_s}{8 \pi N_c}
\left [\ln{\frac{\delta_2}{\delta_{12}}} \ln{\frac{\delta_1}{\delta_{12}}}
+ \ln{\frac{\delta_1}{\delta_{12}}} + \ln{\frac{\delta_2}{\delta_{12}}} + \frac{\pi^2}{3}
\right ]H^{(0)}(p^{\mu}_{2}), \non
G^{(1)}_{3d} &=&   \frac{\alpha_s N_c}{8 \pi} \left [\frac{3}{\epsilon} +
3 \ln{\frac{4 \pi \mu^2}{\delta_{12} Q^2 e^{\gamma_E}}} + \frac{11}{2}\right ]H^{(0)} \non
&& - \frac{\alpha_s N_c}{8 \pi}
\left [ \ln{\frac{\delta_1}{\delta_{12}}} + \ln{\frac{\delta_2}{\delta_{12}}} \right ]H^{(0)}(p^{\mu}_{2})
\non
G^{(1)}_{3e} &=&  \frac{\alpha_s N_c}{8 \pi}
\left [\frac{3}{\epsilon} +
3 \ln{\frac{4 \pi \mu^2}{x_1 Q^2 e^{\gamma_E}}} + \frac{11}{2} \right ]H^{(0)} \non
&& - \frac{\alpha_s N_c}{8 \pi} \left [ \ln{\frac{\delta_2}{x_1}} \ln{x_2}
+\ln{\frac{\delta_2}{x_1}}  \right ]H^{(0)}(x_{1}p^{\mu}_{1}).
\label{eq:vertex1}
\eeq
It is easy to find that the NLO twist-3 corrections to the LO hard kernel
$H^{(0)}$ in Eq.~(\ref{eq:ha0-t3}) have the UV divergence and they have the
same divergence behavior in the self-energy and the vertex corrections.
The summation of these UV divergences leads to the same result as the one
for the NLO twist-2 case \cite{prd83-054029}
\beq
\frac{\alpha_{s}}{4 \pi}\left (11 - \frac{2}{3} N_{f} \right ) \frac{1}{\epsilon},
\label{eq:quarkUV}
\eeq
which meets the requirement of the universality of the wave functions.

The amplitude $G^{(1)}_{3a}$ have no IR divergence due to the fact that
the numerator in the amplitude of the collinear
region is dominated by the transverse contributions which are negligible
in Fig.~\ref{fig:fig3}(a).
IR divergences in $G^{(1)}_{3c}$ and $G^{(1)}_{3d}$ are only relevant with
the hard kernel $H^{(0)}(p^{\mu}_{2})$, which is induced by the singular
gluon attaches to the down quark lines.
Similarly, IR divergences in $G^{(1)}_{3b}$ and $G^{(1)}_{3e}$
only occur in the hard kernel $H^{(0)}(x_{1}p^{\mu}_{1})$
since the singular gluon  is attached to the up quark lines.

The amplitude $G^{(1)}_{3b}$ should have collinear divergence because the radiative
gluon in Fig.~\ref{fig:fig3}(b) is attached to the light valence quark of the final
state pion, and we find that the IR contribution is regulated by $\ln{\delta_{2}}$.
Both the collinear and soft divergences are produced in $G^{(1)}_{3c}$ because the
radiative gluon in Fig.~\ref{fig:fig3}(c) is attached to the external light
valence anti-quarks.
The large double logarithm $\ln{\delta_1} \ln{\delta_2}$ comes
from the overlap  of the IR divergences, and will be canceled by the
large double logarithm term from Fig.~\ref{fig:fig4}(f).

The radiative gluon in Fig.~\ref{fig:fig3}(d) is attached to the light valence
anti-quarks as well as the virtual LO hard gluon, so the soft divergence and
the large double logarithm aren't generated in $G^{(1)}_{3d}$.
The radiative gluon in Fig.~\ref{fig:fig3}(e) is attached only to the light
valence quark as well as the virtual LO hard gluon, and then $G^{(1)}_{3e}$
just contains the collinear divergence regulated by $\ln{\delta_2}$
in the $l \parallel P_2$ region.

%%-----------------------------------------------------------------------
%%-----------------------------------------------------------------------

The NLO twist-3 contributions from the box and pentagon diagrams as shown in
Fig.~\ref{fig:fig4} are summarized as
\beq
&&G^{(1)}_{4a} = - \frac{\alpha_s  N_c}{8 \pi}
\left [(1 + \ln{x_{1}}) \ln{\delta_{1}} -
(1 + \frac{3}{2} \ln{x_{1}}) \ln{x_{2}} + \frac{1}{8} +  \frac{\pi^2}{12} \right ]
H^{(0)}(x_{1}p^{\mu}_{1}), \non
&&~~~~~~~~       - \frac{\alpha_s  N_c}{8 \pi}
\left [(1 + \ln{x_{2}}) \ln{\delta_{2}} -
(1 + \frac{3}{2} \ln{x_{2}}) \ln{x_{1}} + \frac{1}{8} +  \frac{\pi^2}{12}
\right ]H^{(0)}(x_{2}p^{\mu}_{2}), \non
&&G^{(1)}_{4b} \equiv 0, \non
&&G^{(1)}_{4c} = -\frac{\alpha_s}{8 \pi  N_c}
\left [\ln{\frac{x_{1}}{\delta_{2}}} \ln{\frac{x_{2}}{\delta_{1}}}
+ \ln^{2}{x_{2}} \right ]H^{(0)}(x_{1}p^{\mu}_{1}) \non
&&~~~~~~~~       -\frac{\alpha_s}{8 \pi  N_c}
\left [\ln{\frac{x_{2}}{\delta_{1}}} \ln{\frac{x_{1}}{\delta_{2}}}
+ \ln^{2}{x_{1}} \right ]H^{(0)}(x_{2}p^{\mu}_{2}),\non
&&G^{(1)}_{4d} \equiv 0, \non
&&G^{(1)}_{4e} =  \frac{\alpha_s}{8 \pi N_c}
\left [\ln{\delta_{1}} \ln{\delta_{2}} +
\ln{\delta_{1}} + \frac{5}{4} \right ] H^{(0)}(x_{1}p^{\mu}_{1}), \non
&&G^{(1)}_{4f} = -\frac{\alpha_s}{8 \pi N_c}
\left [\ln{\frac{\delta_1}{\delta_{12}}} \ln{\frac{\delta_2}{\delta_{12}}}
- 2 \ln{2} - 1 \right ]H^{(0)}(p^{\mu}_{2}).
\label{eq:boxp1}
\eeq
Because of the properties of the propagators in above four- and five-point integrals,
there is no UV divergences in above amplitudes.
Fig.~\ref{fig:fig4}(b) and 4(d) are two-particle reducible diagrams, their
contribution should be canceled by the corresponding effective diagrams
Fig.~\ref{fig:fig5}(c) and Fig.~\ref{fig:fig6}(c) for the NLO initial and final
state meson wave functions due to the requirement of the factorization theorem,
so we can set them zero safely.

The $H^{(0)}(x_{2}p^{\mu}_{2})$  terms appeared in $G_{4a}^{(1)}$ and $G_{4a}^{(1)}$ are obtained from the
evaluation of Fig.~4(a) and 4(c) only. The LO hard kernel $H^{(0)}(x_{2}p^{\mu}_{2})$ has the same form
as the $H^{(0)}(x_1 p^{\mu}_1)$ as defined in Eq.~(\ref{eq:lot3hka1}) but with
replacements of $x_1\to x_2$ and $p_1^{\mu}\to p_2^\mu$.
IR regulators only appear to the hard kernel $H^{(0)}(x_{1}p^{\mu}_{1})$ of the Fig.~\ref{fig:fig4}(a,c,e), which are
decided by the fact that the left end-point of the emission gluon is attached to the up  light external line.
Similarly, Fig.~\ref{fig:fig4}(f) only grow the IR regulators to the hard kernel $H^{(0)}(p^{\mu}_{2})$.
Note that the emission gluon in Fig.~\ref{fig:fig4}(c,e,f) is attached to external light lines, so it's amplitude
would dominated in the collinear regions and soft region, then the double logarithm would appear.
The attaching of the emission gluon in Fig.~\ref{fig:fig4}(a) to the initial external line and the LO hard kernel
deduce that only IR regulator $\ln{\delta_{1}}$ is grown in the amplitude $G^{(1)}_{4a}$.

Now we just consider the IR parts regulated by $\ln{\delta_{i}}$ which would not be
canceled directly by their counterparts from the effective diagrams of Fig.~\ref{fig:fig5}.
These IR pieces appear in $G^{(1)}_{3b,3c,3d,3e}$ and $G^{(1)}_{4a,4c,4e,4f}$.
We class these amplitudes into two sets according to the hard kernels to which those
IR regulators $\ln{\delta_{i}}$ give corrections.
Then the first set includes $G^{(1)}_{3c,3d}$ and $G^{(1)}_{4f}$, while
the second set contains $G^{(1)}_{3b,3e}$ and $G^{(1)}_{4a,4c,4e}$ terms.
These amplitudes are calculated in the leading IR regions to check the $k_{T}$
factorization theorem.

We firstly evaluate the NLO twist-3 corrections to
$H^{(0)}(p^{\mu}_{2})$. The amplitudes $G^{(1)}_{3c,3d}$ and $G^{(1)}_{4f}$
are recalculated by employing the phase space slicing method \cite{prd65-094032},
\beq
&&G^{(1)}_{3c}( l\to 0) = \frac{\alpha_{s}}{8 \pi N_{c}}
\left [\ln{\frac{\delta_{1}}{\delta_{12}}}
\ln{\frac{\delta_{2}}{\delta_{12}}} + \frac{\pi^{2}}{3}
\right ] H^{(0)}(p^{\mu}_{2}), \non
&&G^{(1)}_{3c}(l \parallel p_{1}) = \frac{\alpha_{s}}{8 \pi N_{c}}
\left [\ln{\frac{\delta_{1}}{\delta_{12}}}
\ln{\frac{\delta_{2}}{\delta_{12}}} + \ln{\frac{\delta_{1}}{\delta_{12}}}
\right ] H^{(0)}(p^{\mu}_{2}), \non
&&G^{(1)}_{3c}(l \parallel p_{2}) = \frac{\alpha_{s}}{8 \pi N_{c}}
\left [\ln{\frac{\delta_{1}}{\delta_{12}}}
\ln{\frac{\delta_{2}}{\delta_{12}}} + \ln{\frac{\delta_{2}}{\delta_{12}}}
\right ] H^{(0)}(p^{\mu}_{2}). \label{eq:g3c1}
\eeq
\beq
&&G^{(1)}_{3d}(l \parallel p_{1}) = \frac{\alpha_{s} N_{c}}{8 \pi}
\left [- \ln{\frac{\delta_{1}}{\delta_{12}}} \right ] H^{(0)}(p^{\mu}_{2}), \non
&&G^{(1)}_{3d}(l \parallel p_{2}) = \frac{\alpha_{s} N_{c}}{8 \pi}
\left [- \ln{\frac{\delta_{2}}{\delta_{12}}} \right ] H^{(0)}(p^{\mu}_{2}).
\label{eq:g3d1}
\eeq
\beq
&&G^{(1)}_{4f}(l\to 0) = - \frac{\alpha_{s}}{8 \pi N_{c}}
\left [\ln{\frac{\delta_{1}}{\delta_{12}}}
\ln{\frac{\delta_{2}}{\delta_{12}}} + \frac{\pi^{2}}{3}
\right ] H^{(0)}(p^{\mu}_{2}), \non
&&G^{(1)}_{4f}(l \parallel p_{1}) = - \frac{\alpha_{s}}{8 \pi N_{c}}
\left [\ln{\frac{\delta_{1}}{\delta_{12}}}
\ln{\frac{\delta_{2}}{\delta_{12}}} + \frac{\pi^{2}}{6} - 1
\right ] H^{(0)}(p^{\mu}_{2}), \non
&&G^{(1)}_{4f}(l \parallel p_{2}) = - \frac{\alpha_{s}}{8 \pi N_{c}}
\left [\ln{\frac{\delta_{1}}{\delta_{12}}}
\ln{\frac{\delta_{2}}{\delta_{12}}} + \frac{\pi^{2}}{6} -2 \ln{2}
\right ] H^{(0)}(p^{\mu}_{2}).
\label{eq:g4d1}
\eeq
By summing up all terms in Eq.~(\ref{eq:g3c1},\ref{eq:g3d1},\ref{eq:g4d1}),
one finds that the soft contributions in the limit $l\to 0$
from Fig.~\ref{fig:fig3}(c) and Fig.~\ref{fig:fig4}(f) are canceled each other,
while the remaining collinear contributions in the regions of $l \parallel p_{1}$
and $l \parallel p_2$ are of the form,
\beq
G^{(1)}_{3c+3d+4f}(l \parallel p_{1}) &=& - \frac{\alpha_s C_F}{8 \pi}
\left [2 \ln{\delta_{1}} \right ] H^{(0)}(p^{\mu}_{2}), \\
G^{(1)}_{3c+3d+4f}(l \parallel p_{2}) &=& - \frac{\alpha_s C_F}{8 \pi}
\left [2 \ln{\delta_{2}} \right ] H^{(0)}(p^{\mu}_{2}),
\label{eq:collp2}
\eeq

The IR contributions to NLO twist-3 corrections to $H^{(0)}(x_{1}p^{\mu}_{1})$
can be obtained in similar way.
\beq
G^{(1)}_{3b}(l \parallel p_{2}) &=& - \frac{\alpha_{s}}{8 \pi N_{c}}
\left [ 1 - \ln{\frac{\delta_{2}}{x_{1}}} \right ] H^{(0)}(x_{1}p^{\mu}_{1}), \label{eq:g3b1}\\
G^{(1)}_{3e}(l \parallel p_{2}) &=& - \frac{\alpha_{s} N_{c}}{8 \pi}
\left[\ln{\frac{\delta_{2}}{x_{1}}}
(\ln{x_{2} + 1)} \right ] H^{(0)}(x_{1}p^{\mu}_{1}), \label{eq:g3e1}\\
G^{(1)}_{4a}(l \parallel p_{1}) &=& - \frac{\alpha_{s} N_{c}}{8 \pi}
\left [\ln{\delta_{1}}(\ln{x_{1}} + 1)
- \ln{x_{2}}(\frac{3}{2}\ln{x_{1}} + 1) \right.\non
&& \left. + \frac{\pi^2}{12} + \frac{1}{8} \right ] H^{(0)}(x_{1}p^{\mu}_{1}),
\label{eq:g4a1}
\eeq
\beq
G^{(1)}_{4c}(l\to 0) &=& - \frac{\alpha_{s}}{8 \pi N_{c}}
\left [\ln{\delta_{1}} \ln{\delta_{2}} +
\frac{\pi^2}{3} \right ] H^{(0)}(x_{1}p^{\mu}_{1}), \non
G^{(1)}_{4c}(l \parallel p_{1}) &=&  - \frac{\alpha_{s}}{8 \pi N_{c}}
\left[\ln{\delta_{1}} \ln{\frac{\delta_{2}}{x_{1}}}
+ \frac{\pi^2}{6} \right ] H^{(0)}(x_{1}p^{\mu}_{1}), \non
G^{(1)}_{4c}(l \parallel p_{2}) &=&  - \frac{\alpha_{s}}{8 \pi N_{c}}
\left[\ln{\delta_{2}} \ln{\frac{\delta_{1}}{x_{2}}}
+ \ln{x_{2}}(\ln{x_{2}} + \ln{x_{1}}) + \frac{\pi^2}{6} \right ] H^{(0)}(x_{1}p^{\mu}_{1}),
\label{eq:g4c1}
\eeq
\beq
G^{(1)}_{4e}(l\to 0) &=& \frac{\alpha_{s}}{8 \pi N_{c}}
\left [\ln{\delta_{1}} \ln{\delta_{2}} +
\frac{\pi^2}{3} \right ] H^{(0)}(x_{1}p^{\mu}_{1}), \non
G^{(1)}_{4e}(l \parallel p_{1}) &=& \frac{\alpha_{s}}{8 \pi N_{c}}
\left [\ln{\delta_{1}} \ln{\delta_{2}} + \ln{\delta_{1}} - \ln{x_{1}}
+ \frac{\pi^2}{6} + 3 \right ] H^{(0)}(x_{1}p^{\mu}_{1}), \non
G^{(1)}_{4e}(l \parallel p_{2}) &=& \frac{\alpha_{s}}{8 \pi N_{c}}
\left [\ln{\delta_{1}} \ln{\delta_{2}} + \frac{3}{2} \ln{x_{1}}
+ \frac{\pi^2}{6} - \frac{7}{4} \right] H^{(0)}(x_{1}p^{\mu}_{1}).
\label{eq:g4e1}
\eeq
Again, the soft parts from Fig.~\ref{fig:fig4}(c) and Fig.~\ref{fig:fig4}(e)
are canceled each other, while the remaining collinear contributions
to the LO hard kernel $H^{(0)}(x_{1}p^{\mu}_{1})$,
after summing up the amplitudes as given in Eqs.~(\ref{eq:g3b1},\ref{eq:g3e1},\ref{eq:g4a1},
\ref{eq:g4c1},\ref{eq:g4e1}), are the following
\beq
G^{(1)}_{4a+4c+4e}(l \parallel p_{1}) &=&  - \frac{\alpha_{s} C_F}{8 \pi}
\left [2 \ln{\delta_{1}}(\ln{x_{1}} + 1)\right ] H^{(0)}(x_{1}p^{\mu}_{1}), \non
G^{(1)}_{3b+3e+4c+4e}(l \parallel p_{2}) &=&  - \frac{\alpha_{s} C_F}{8 \pi}
\left [2 \ln{\delta_{2}}(\ln{x_{2}} + 1)\right ] H^{(0)}(x_{1}p^{\mu}_{1}).
\label{eq:collx1p1}
\eeq
Note that we have dropped the constant terms in Eqs.~(\ref{eq:collp2},\ref{eq:collx1p1})
since we here consider the IR parts only.
According to previous studies in Refs.\cite{prd76-034008,prd83-054029,prd85-074004},
we know that these IR divergences could be absorbed into the NLO wave functions
of the pion mesons.  This point will become clear after we complete the
calculations for the effective diagrams in Fig.~\ref{fig:fig5} and Fig.~\ref{fig:fig6}.
This absorption means that the $k_{T}$ factorization is valid at the NLO level
for the $\pi \gamma^* \to \pi$ process.

Without the reducible diagrams $G^{(1)}_{2a,2b,2c,2d,4b,4d}$,
the summation for the NLO twist-3 contributions from all the irreducible QCD
quark diagrams as illustrated by Figs.~(\ref{fig:fig2},\ref{fig:fig3},\ref{fig:fig4})
leads to the final result for $G^{(1)}$:
\beq
G^{(1)} &=& \frac{\alpha_s C_F}{8 \pi} \Biggl [ \frac{29}{2}
\left ( \frac{1}{\epsilon} + \ln{\frac{4 \pi \mu^2}{Q^2 e^{\gamma_E}}} \right )
- 2 \ln{\delta_{1}(\ln{x_{1}} + 1) -2 \ln{\delta_{2}}}(\ln{x_{2}} + 1) \non
&&- \frac{21}{8} \ln{(x_{1}x_{2})} - \frac{23}{8} \ln{x_{1}} - \frac{1}{4}\ln^{2}{x_{2}} - \frac{9}{4}\ln{x_{2}}
- \frac{3 \pi^{2}}{16} + \frac{721}{32} \Biggr ] H^{(0)}(x_{1}p^{\mu}_{1}) \non
&&+\frac{\alpha_s C_F}{8 \pi} \Biggl [ \frac{29}{2}
\left (\frac{1}{\epsilon} + \ln{\frac{4 \pi \mu^2}{Q^2 e^{\gamma_E}}} \right )
- 2 \ln{\delta_{1} -2 \ln{\delta_{2}}} \non
&&- 4 \ln{(x_{1}x_{2})} - 5 \ln{x_{1}} + \frac{\pi^{2}}{12} + \frac{\ln{2}}{2} +23
\Biggr ] H^{(0)}(p^{\mu}_{2}),
\label{eq:nloqd}
\eeq
for $N_f = 6$. The UV divergence in above expression is the same one as in the pion
electromagnetic form factor \cite{prd83-054029}, which determines the
renormalization-group(RG) evolution of the strong coupling constant $\alpha_s$.

\subsection{Convolution of the $O(\alpha_s)$ wave functions with the LO hard kernel}

A basic argument of $k_T$ factorization is that the IR divergences
from the NLO corrections can also be absorbed into the non-perturbative wave
functions which are universal.
From this point, the convolution of the NLO wave functions and the LO hard kernel
$H^{(0)}$ are computed, and the resultant IR part should cancel the IR divergences
appeared in the NLO amplitude $G^{(1)}$ as given in Eq.~(\ref{eq:nloqd}).

The convolution of the NLO pion wave functions and the LO hard kernel are calculated
in this subsection.
In $k_T$ factorization theorem, the $\Phi^{(1)}_{\pi,P}$ and $\Phi^{(1)}_{\pi,T}$
collect the $O(\alpha_s)$ effective diagrams for the twist-3 transverse momentum
dependent (TMD) light-cone wave function $\Phi_{\pi,P}$
and $\Phi_{\pi,T}$ respectively\cite{prd64-014019,epjc40-395}.
In the $\pi \gamma^{\star} \to \pi$ process we calculate, only the $O(\alpha_s)$ order
pseudoscalar component $\Phi^{(1)}_{\pi,P}$ of the final state pion,
but both the $\Phi^{(1)}_{\pi,P}$ and $\Phi^{(1)}_{\pi,T}$ components
of the initial pion should be convoluted with  the LO hard kernel.
\beq
\Phi_{\pi,P}(x'_1,k'_{1T};x_1,k_{1T}) &=&
\int \frac{d y^-}{2 \pi} \frac{d^2 y_T}{(2 \pi)^2} e^{-i x'_1 P^+_1 y^-
+ i \textbf{k}'_{1T} \cdot \textbf{y}_T} \non
&& \hspace{-2cm}\cdot <0\mid\overline{q}(y) \gamma_5 W_y(n_1)^{\dag}
I_{n_1;y,0} W_0(n_1) q(0) \mid \overline{u}(P_1 - k_1) d(k_1)>,
\label{eq:phi11}\\
\Phi_{\pi,T}(x'_1,k'_{1T};x_1,k_{1T}) &=&
\int \frac{d y^-}{2 \pi} \frac{d^2 y_T}{(2 \pi)^2} e^{-i x'_1 P^+_1 y^-
+ i \textbf{k}'_{1T} \cdot \textbf{y}_T} \non
&& \hspace{-2cm}\cdot <0\mid\overline{q}(y) \gamma_5 (\nsl_{+}\nsl_{-}
-1 ) W_y(n_1)^{\dag}
I_{n_1;y,0} W_0(n_1) q(0) \mid \overline{u}(P_1 - k_1) d(k_1)>,
\label{eq:nlowfpion1pt}\\
\Phi_{\pi,P}(x_2,k_{2T};x'_2,k'_{2T}) &=& \int \frac{d z^+}{2 \pi}
\frac{d^2 z_T}{(2 \pi)^2} e^{-i x'_2 P^-_2 z^+ + i \textbf{k}'_{2T}
\cdot \textbf{z}_T} \non
&&\hspace{-2cm}\cdot <0\mid\overline{q}(z) W_z(n_2)^{\dag} I_{n_2;z,0}
W_0(n_2) \gamma_5 q(0)  \mid u(P_2 - k_2) \overline{d}(k_2)>,
\label{eq:nlowfpion2p}
\eeq
where $y = (0, y^-, \textbf{y}_T)$ and $z = (z_+, 0, \textbf{z}_T)$
are the light-cone coordinates of the anti-quark field $\bar{q}$
carrying the momentum faction $x_i$, respectively.
The Wilson lines with the choice of $n^2_i \neq 0$ to avoid the light-cone
singularity \cite{prd85-074004,jhep0601-067} are defined as
\beq
W_y(n_1) &=& {\rm P}\; \exp [-i g_s \int^{\infty}_{0} d \lambda n_1 \cdot A(y + \lambda n_1)],
\label{eq:wl01}\\
W_z(n_2) &=& {\rm P}\; \exp [-i g_s \int^{\infty}_{0} d \lambda n_1 \cdot A(z + \lambda n_2)],
\label{eq:wl02}
\eeq
where P is the path ordering operator.
The two Wilson line $W_y(n_i)$ ( $W_z(n_i)$) and $W_{0}(n_i)$ are connected by a
vertical link $I_{n_i;y,0}$ ($I_{n_i;z,0}$ ) at infinity \cite{plb543-66}.
Then the additional light-cone singularities from the region where loop
momentum $l\parallel n_-(n_+) $ \cite{appb34-3103} are regulated by the IR regulator
$n^2_1\neq 0$ and $n^2_2\neq 0$.
The scales $\xi^2_1 \equiv 4 (n_1 \cdot p_1)^2/ |n^2_1| = Q^2 |n^-_1/n^+_1|$ and $\xi^2_2 \equiv 4
(n_2 \cdot p_2)^2/ |n^2_2| = Q^2 |n^+_2/n^-_2|$ are introduced to
decide the wave functions of the initial and final state pion respectively.
It is important to emphasize that the variation of the above scales can be treated
as a factorization scheme-dependence, which entered the hard kernel when taking
the difference of the quark diagrams in full QCD
and the effective diagrams for the wave functions in NLO calculations.
Recently, the above scheme-dependent rapidity logarithms were diminished
by joint resummation\cite{arxiv1308-0413} for $B$ meson wave functions\cite{jhep1302-008},
and for pion wave function and pion transition form factor\cite{arxiv1310-3672}.
In this paper we minimize the above scheme-dependent scales by adhering them to fixed
$n^2_1$ and $n^2_2$.

%%-----------------------------------------------------------------------
\begin{figure*}
\centering
\vspace{-2cm}
\leftline{\epsfxsize=12cm\epsffile{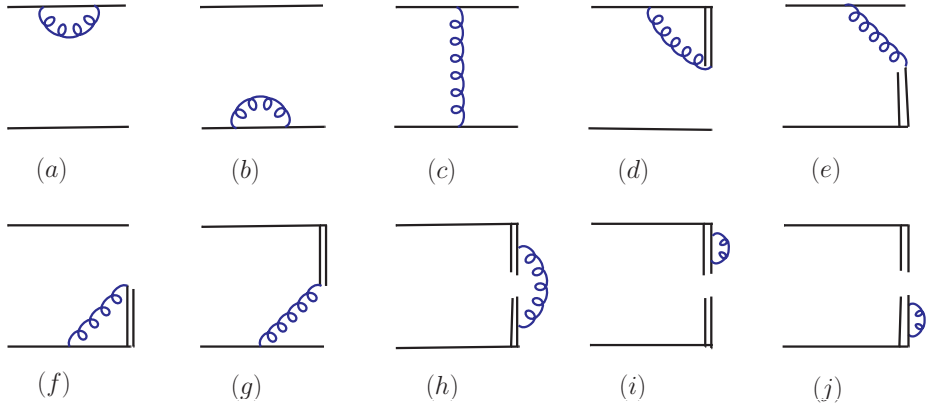}}
\vspace{-10cm}
\caption{The effective $O(\alpha_s)$ diagrams for the initial $\pi$ meson
wave function.}
\label{fig:fig5}
\end{figure*}
%%-----------------------------------------------------------------------

The convolution of the $O(\alpha_s)$ initial state wave function as shown in
Fig.~\ref{fig:fig5}  and $H^{(0)}$ over the integration variables $x'_1$ and
$k'_{1T}$ is of the form
\beq
\Phi^{(1)}_{\pi} \otimes H^{(0)} \equiv \int dx'_1 d^2 \textbf{k}'_{1T} \Phi^{(1)}_{\pi}(x_1,\textbf{k}_{1T};
x'_1,\textbf{k}'_{1T}) H^{(0)}(x'_1,\textbf{k}'_{1T};x_2,\textbf{k}_{2T}).
\label{eq:convob}
\eeq
When making this convolution, the $n_1$ is chosen approximately as the
vector $n_-$ with a very small plus component $n^+_1$ to avoid the light-cone
singularity. Note that the sign of $n^-_1$ is positive while the
sign of $n^+_1$ can be positive or negative for convenience.
The results after making the convolution for each figure in Fig.~\ref{fig:fig5}
are given in the following  with $\mu_f$ being the factorization scale.
\beq
&&\Phi^{(1)}_{5a} \otimes H^{(0)} = -\frac{\alpha_s C_F}{8 \pi}
\left [\frac{1}{\epsilon} +
\ln{\frac{4 \pi \mu^2_f}{\delta_{1} Q^2 e^{\gamma_E}}} + 2 \right ] H^{(0)}, \non
&&\Phi^{(1)}_{5b} \otimes H^{(0)} = -\frac{\alpha_s C_F}{8 \pi}
\left [\frac{1}{\epsilon} +
\ln{\frac{4 \pi \mu^2_f}{\delta_{1} Q^2 e^{\gamma_E}}} + 2\right ] H^{(0)}, \non
&&\Phi^{(1)}_{5c} \otimes H^{(0)} \equiv 0, \non
&&\Phi^{(1)}_{5d} \otimes H^{(0)} = \frac{\alpha_s C_F}{8 \pi} \left [\frac{1}{\epsilon} +
\ln{\frac{4 \pi \mu^2_f}{\xi^{2}_{1} e^{\gamma_E}}} - \ln^{2}{(\delta_{1}r_{Q})}
- 2 \ln{(\delta_{1} r_Q)} - \frac{ \pi^{2}}{3} +2 \right ] H^{(0)}(x_{1}p^{\mu}_{1}), \non
&&\Phi^{(1)}_{5e} \otimes H^{(0)} = \frac{\alpha_s C_F}{8 \pi}
\left [\ln^2{(\frac{\delta_{1} r_Q}{x_{1}})}
+ \pi^{2} \right ] H^{(0)}(x_{1}p^{\mu}_{1}), \non
&&\Phi^{(1)}_{5f} \otimes H^{(0)} = \frac{\alpha_s C_F}{8 \pi}
\left [\frac{1}{\epsilon} +
\ln{\frac{4 \pi \mu^2_f}{\xi^{2}_{1} e^{\gamma_E}}}
- \ln^{2}{(\frac{\delta_{1} r_{Q}}{x^{2}_{1}})}
- 2 \ln{(\frac{\delta_{1} r_{Q}}{x^{2}_{1}})} - \frac{ \pi^{2}}{3} +2 \right ] H^{(0)}(p^{\mu}_{2}), \non
&&\Phi^{(1)}_{5g} \otimes H^{(0)} = \frac{\alpha_s C_F}{8 \pi}
\left [\ln^{2}{(\frac{\delta_{1} r_{Q}}{x^{2}_{1}})}
- \frac{\pi^{2}}{3} \right ] H^{(0)}(p^{\mu}_{2}), \non
&&(\Phi^{(1)}_{5h} + \Phi^{(1)}_{5i} + \Phi^{(1)}_{5j}) \otimes H^{(0)} = \frac{\alpha_s
C_F}{4 \pi} \left [\frac{1}{\epsilon} +
\ln{\frac{4 \pi \mu^2_f}{Q^2 e^{\gamma_E}}} - \ln{\delta_{12}} \right ] H^{(0)}.
\label{eq:nloedi}
\eeq
The dimensionless parameter $r_{Q} = Q^2 / \xi^2_1$ is defined to simplify
the expressions as given in above equation.
In Eq.~(\ref{eq:nloedi}) all the IR divergence are regulated by $\ln{\delta_{1}}$ in
the convolution $\Phi^{(1)}_{\pi} \otimes H^{(0)}$.
Fig.~\ref{fig:fig5}(d,e) just give the corrections to the LO hard kernel $H^{(0)}(x_{1}p^{\mu}_{1})$,
while Fig.~\ref{fig:fig5}(f,g) provide the corrections to the LO hard kernel $H^{(0)}(p^{\mu}_{2})$,
because the gluon attaches to the Wilson line and the up external line in the former two subdiagrams
( Fig.~\ref{fig:fig5}(d) and 5(e) ), but attaches to the Wilson line and the down external line in the later two
subdiagrams ( Fig.~\ref{fig:fig5}(f) and 5(g) ).
The corrections from Fig.~\ref{fig:fig5}(a,b) are canceled by those from Fig.~\ref{fig:fig2}(a,b).
It is unnecessary to calculate the reducible subdiagram Fig.~\ref{fig:fig5}(c), since it will be
canceled by Fig.~\ref{fig:fig4}(b) completely.

Only the three-point integral was involved in the convolution of
$\Phi^{(1)}_{5d} \otimes H^{(0)}$ and
$\Phi^{(1)}_{5f} \otimes H^{(0)}$, because there is no momenta flow
into the LO hard kernel in these two sundiagrams.
A four-point integral should been calculated in the convolution $\Phi^{(1)}_{5e} \otimes H^{(0)}$
because  the momenta is flow into the LO hard kernel and the $H^{(0)}(x_{1}p^{\mu}_{1})$ cancels an denominator.
The convolution $\Phi^{(1)}_{5g} \otimes H^{(0)}$ involves a five-point integral due to the flow momenta and the
correction to the LO hard kernel $H^{(0)}(p^{\mu}_{2})$.
After summing up all  $O(\alpha_s)$ contributions in Fig.~\ref{fig:fig5} except those from the reducible subdiagrams
Fig.~\ref{fig:fig5}(a), 5(b) and 5(c), we find
\beq
\Phi^{(1)}_{\pi} \otimes H^{(0)} &=& \frac{\alpha_s C_F}{8 \pi}
\Bigl [ \frac{3}{\epsilon} + 3 \ln{\frac{4 \pi}{e^{\gamma_E}}} + 3 \ln{\frac{\mu^2_f}{Q^{2}}}
- 2 \ln{(\delta_1 r_{Q})}(\ln{x_{1}} + 1) \non
&& - 2 \ln{(x_{1} x_{2} r_{Q})} + \frac{2\pi^{2}}{3} - 2 \Bigr ] H^{(0)}(x_{1}p^{\mu}_{1}) \non
&&+ \frac{\alpha_s C_F}{8 \pi}
\Bigl [\frac{3}{\epsilon} + 3 \ln{\frac{4 \pi}{e^{\gamma_E}}}
+ 3 \ln{\frac{\mu^2_f}{Q^{2}}} - 2 \ln{(\delta_1 r_{Q})} \non
&&- 2 \ln{(x_{1} x_{2} r_{Q})} + 4 \ln{x_{1}} - \frac{2\pi^{2}}{3} - 2 \Bigr ] H^{(0)}(p^{\mu}_{2}),
\label{eq:nloedti}
\eeq
where $r_{Q} = Q^2 / \xi^2_1$.

The convolution of the LO hard kernel $H^{(0)}$ and the NLO outgoing pion meson function
$\Phi^{(1)}_{\pi}$ over the integration variables $x'_2$ and $k'_{2T}$ is
\beq
H^{(0)} \otimes \Phi^{(1)}_{\pi} \equiv \int dx'_2 d^2 \textbf{k}'_{2T}
H^{(0)}(x_1,\textbf{k}_{1T};x'_2,\textbf{k}'_{2T}) \Phi^{(1)}_{\pi,P}(x'_2,\textbf{k}'_{2T};x_2,\textbf{k}_{2T}) .
\label{eq:convopi}
\eeq
The unit vector $n_2$ is chosen approximately  as  $n_+$ with a very small minus component $n^-_2$ to
avoid the light-cone singularity in the convolution.
Note that the sign of $n^+_2$ is positive as $P^+_1$ while the sign of $n^-_2$ is arbitrary for convenience.

\begin{figure*}
\centering
\vspace{-2cm}
\leftline{\epsfxsize=12cm\epsffile{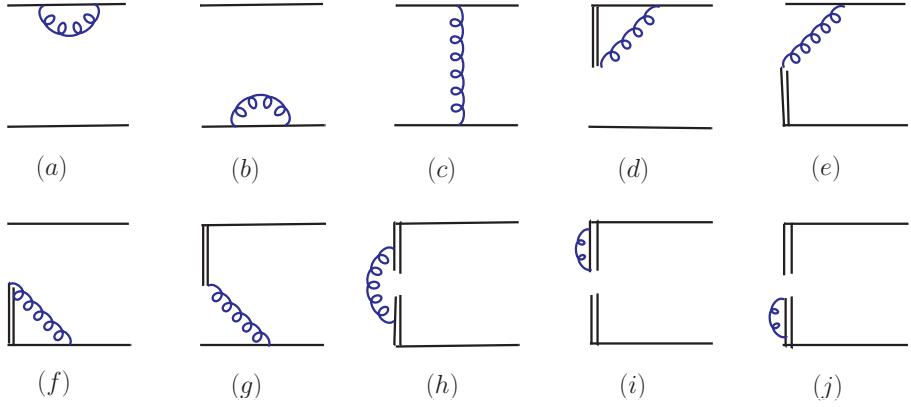}}
\vspace{-10cm}
\caption{The $O(\alpha_s)$ subdiagrams for the final $\pi$ meson wave function.}
\label{fig:fig6}
\end{figure*}

In Fig.~\ref{fig:fig6} we draw all subdiagrams which provide ${\cal O}(\alpha_s)$ NLO corrections to the outgoing pion
wave functions.
Analogous to the case of Fig.~\ref{fig:fig5}, we here make the same evolutions for all subdiagrams in Fig.~\ref{fig:fig6}.
The analytical results for each sunbdiagram of Fig.~\ref{fig:fig6} are listed in the following
with $\mu_f$ being the factorization scale.
\beq
&&H^{(0)} \otimes \Phi^{(1)}_{6a} =-\frac{\alpha_s C_F}{8 \pi} \left [\frac{1}{\epsilon} +
\ln{\frac{4 \pi \mu^2_f}{\delta_{2} Q^2 e^{\gamma_E}}} + 2 \right ] H^{(0)}, \non
&&H^{(0)} \otimes \Phi^{(1)}_{6b} = -\frac{\alpha_s C_F}{8 \pi}  \left[\frac{1}{\epsilon} +
\ln{\frac{4 \pi \mu^2_f}{\delta_{2} Q^2 e^{\gamma_E}}} + 2\right] H^{(0)}, \non
&&H^{(0)} \otimes \Phi^{(1)}_{6c} \equiv 0, \non
&&H^{(0)} \otimes \Phi^{(1)}_{6d} =  \frac{\alpha_s C_F}{8 \pi}  \left[\frac{1}{\epsilon} +
\ln{\frac{4 \pi \mu^2_f}{\xi^{2}_{2} e^{\gamma_E}}} - \ln^{2}{(\delta_{2}\gamma_{Q})}
- 2 \ln{(\delta_{2}\gamma_{Q})} - \frac{ \pi^{2}}{3} +2\right] H^{(0)}(x_{1}p^{\mu}_{1}),
\label{eq:nloedtf0}
\eeq
\beq
&&H^{(0)} \otimes \Phi^{(1)}_{6e} =  \frac{\alpha_s C_F}{8 \pi}  \left[\ln^2{(\frac{\delta_{2}
r_Q}{x_{1}})} + \pi^{2}\right] H^{(0)}(x_{1}p^{\mu}_{1}), \non
&&H^{(0)} \otimes \Phi^{(1)}_{6f} =  \frac{\alpha_s C_F}{8 \pi}  \left[\frac{1}{\epsilon} +
\ln{\frac{4 \pi \mu^2_f}{\xi^{2}_{2} e^{\gamma_E}}} - \ln^{2}{(\frac{\delta_{2}r_Q}{x^{2}_{2}})}
- 2 \ln{(\frac{\delta_{2}r_Q}{x^{2}_{2}})} - \frac{ \pi^{2}}{3} +2\right] H^{(0)}(p^{\mu}_{2}), \non
&&H^{(0)} \otimes \Phi^{(1)}_{6g} =  \frac{\alpha_s C_F}{8 \pi}  \left[\ln^{2}{(\frac{\delta_{2}r_Q}{x^{2}_{2}})}
- \frac{\pi^{2}}{3}\right] H^{(0)}(p^{\mu}_{2}), \non
&&H^{(0)} \otimes (\Phi^{(1)}_{6h} + \Phi^{(1)}_{6i} + \Phi^{(1)}_{6j}) =  \frac{\alpha_s C_F}{4 \pi} \left [\frac{1}{\epsilon} +
\ln{\frac{4 \pi \mu^2_f}{Q^2 e^{\gamma_E}}} - \ln{\delta_{12}}\right] H^{(0)},
\label{eq:nloedtf}
\eeq
where $r_{Q} = Q^2 / \xi^2_1$.
The most complex integral involved in our calculation for Fig.~\ref{fig:fig6} is the four-point
integral, since the relevant momentum fraction $x'_{2}$ is only appeared in one propagator in
the LO hard kernel $H^{(0)}$.

%%-----------------------------------------------------------------------
%%-----------------------------------------------------------------------

The total contributions from the convolution of the LO hard kernel $H^{(0)}$ and the NLO final pion
meson wave function is obtained by summing up all contributions as listed in above equation,
and we dropped the contributions from those reducible subdiagrams $G^{(1)}(a,b,c)$. The summation from
all irreducible subdiagrams of Fig.~6 leads to the final result:
\beq
H^{(0)} \otimes \Phi^{(1)}_{\pi} &=& \frac{\alpha_s C_F}{8 \pi}
\Bigl [ \frac{3}{\epsilon} + 3 \ln{\frac{4 \pi}{e^{\gamma_E}}} + 3 \ln{\frac{\mu^2_f}{Q^{2}}}
- 2 \ln{(\delta_2 r_Q)}(\ln{x_{2}} + 1) \non
&& - 2 \ln{(x_{1} x_{2} r_Q)} + \frac{2\pi^{2}}{3} - 2 \Bigr ] H^{(0)}(x_{1}p^{\mu}_{1}) \non
&& + \frac{\alpha_s C_F}{8 \pi}
\Bigl [\frac{3}{\epsilon} + 3 \ln{\frac{4 \pi}{e^{\gamma_E}}} + 3 \ln{\frac{\mu^2_f}{Q^{2}}} - 2 \ln{(\delta_2 r_Q)} \non
&& - 2 \ln{(x_{1} x_{2} r_Q)} + 4 \ln{x_{2}} - \frac{2\pi^{2}}{3} - 2 \Bigr ] H^{(0)}(p^{\mu}_{2}),
\label{eq:nloed}
\eeq
where $r_{Q} = Q^2 / \xi^2_1$.

\subsection{The NLO twist-3 Hard Kernel }

The IR-finite $k_T$ dependent NLO twist-3 hard kernel for the $\pi \gamma^* \to \pi$ form factor
is derived by taking the difference between the contributions from the quark diagrams in full QCD
and the contributions from the effective diagrams for pion wave functions \cite{prd67-034001}.
\beq
&&H^{(1)}(x_1,\textbf{k}_{1T};x_2,\textbf{k}_{2T}) = G^{(1)}(x_1,\textbf{k}_{1T};x_2,\textbf{k}_{2T}) \non
&&~~~~~~~~~~~~~~- \int dx'_1 d^2 \textbf{k}'_{1T} \Phi^{(1)}_{\pi}(x_1,\textbf{k}_{1T};x'_1,\textbf{k}'_{1T})
H^{(0)}(x'_1,\textbf{k}'_{1T};x_2,\textbf{k}_{2T}) \non
&&~~~~~~~~~~~~~~- \int dx'_2 d^2 \textbf{k}'_{2T} H^{(0)}(x_1,\textbf{k}_{1T};x'_2,\textbf{k}'_{2T})
\Phi^{(1)}_{\pi,P}(x'_2,\textbf{k}'_{2T};x_2,\textbf{k}_{2T}),
\label{eq:nlohk}
\eeq
where $\Phi^{(1)}_{\pi}(x_1,\textbf{k}_{1T};x'_1,\textbf{k}'_{1T})$ include two parts:
$\Phi^{(1)}_{\pi,P}(x_1,\textbf{k}_{1T};x'_1,\textbf{k}'_{1T})$ and
$\Phi^{(1)}_{\pi,T}(x_1,\textbf{k}_{1T};x'_1,\textbf{k}'_{1T})$.

The bare coupling constant $\alpha_s$ in Eq.~(\ref{eq:nloqd},\ref{eq:nloedti},\ref{eq:nloed})
can be rewritten as
\beq
&&\alpha_s = \alpha_s(\mu_f) + \delta Z(\mu_f) \alpha_s(\mu_f),
\label{eq:renormcc}
\eeq
in which the counterterm $\delta Z(\mu_f)$ is defined in the modified minimal subtraction scheme. Inserting
Eq.~(\ref{eq:renormcc}) into Eq.~(\ref{eq:lot3hka},\ref{eq:nloqd},\ref{eq:nloedti},\ref{eq:nloed})
regularizes the UV poles in Eq.~(\ref{eq:nlohk}) through the term
$\delta Z (\mu_f) H^{(0)}$, and then the UV poles in Eq.~(\ref{eq:nloedti},\ref{eq:nloed})
are regulated by the counterterm of the quark field and by an additional
counterterm in the modified minimal subtraction scheme.
The UV behavior of the NLO twist-3 contributions is the same as the NLO twist-2 ones, which satisfy the requirement
of the universality of the pion wave functions.

Based on the above calculations, it is straightforward to write down the NLO twist-3 hard kernel $H^{(1)}$
for Fig.~\ref{fig:fig1}(a), assuming $\xi^{2}_{1} \equiv \xi^{2}_{2} \equiv Q^2$:
\beq
H^{(1)} &=& \frac{\alpha_s(\mu_f) C_F}{8 \pi} \Biggl[ \frac{21}{2} \ln{\frac{\mu^2}{Q^2}} - 6 \ln{\frac{\mu^2_f}{Q^2}}
- \frac{53}{8} \ln{(x_1 x_2)} \non
&&- \frac{23}{8} \ln{x_1} - \frac{4}{9} \ln{x_2} - \frac{1}{4} \ln^{2}{x_2} - \frac{137}{48} \pi^{2}
+ \frac{337}{32} \Biggr] H^{(0)}(x_{1}p^{\mu}_{1}) \non
&&  + \frac{\alpha_s(\mu_f) C_F}{8 \pi} \Biggl [ \frac{21}{2} \ln{\frac{\mu^2}{Q^2}} - 6 \ln{\frac{\mu^2_f}{Q^2}}
- 8 \ln{(x_1 x_2)}\non
&& -  \ln{x_1} + 4 \ln{x_2} - \frac{31}{12} \pi^{2} + \frac{1}{2} \ln{2} + 11 \Biggr ] H^{(0)}(p^{\mu}_{2}),
\label{eq:nlohk1}
\eeq
where $\mu$ and $\mu_f$ are the renormalization scale and factorization scale respectively.
Following the schemes in the NLO analysis of the $B \to \pi$ transition form factor at the leading twist
\cite{prd85-074004}, we here also set $\xi^2_2=Q^2$ in order to to obtain a simple expression as
given in Eq.~(\ref{eq:nlohk1}).

The additional double logarithm $\ln^{2}{x_1}$, derived from the limit that the internal quark is
on-shell due to the tiny momentum fraction $x_1$,  should also be considered.
It can be absorbed into the Jet function $J(x_1)$ as in Refs.~\cite{prd66-094010,plb555-197}
\beq
J^{(1)} H^{(0)} &=& - \frac{1}{2} \frac{\alpha_s(\mu_f) C_F}{4 \pi}
\left [\ln^2{x_1} + \ln{x_1} + \frac{\pi^2}{3} \right ] H^{(0)}(p^{\mu}_2),
\label{eq:jetfunction}
\eeq
where the factor $\frac{1}{2}$ reflects the different spin structure of the twist-3 and twist-2 parts.
There exists no Jet function $J(x_2)$ because the momentum fraction $x_{2}$ wouldn't grow end-point singularity.
The NLO twist-3 hard kernel $H^{(1)}$ in Eq.~(\ref{eq:nlohk1}) will become the following  form
after subtracting out the Jet function in Eq.~(\ref{eq:jetfunction})
\beq
H^{(1)}(x_i,\mu,\mu_f,Q^2) &\to& H^{(1)} - J^{(1)} H^{(0)} \non
&\equiv& F^{(1)}_{\rm T3,A1}(x_i,\mu,\mu_f,Q^2) H^{(0)}(x_{1} p^{\mu}_{1})
+ F^{(1)}_{\rm T3,A2}(x_i,\mu,\mu_f,Q^2) H^{(0)}(p^{\mu}_{2}),
\label{eq:nlohk2}
\eeq
where the two factors of the NLO twist-3 contributions for Fig.1(a) are of the form
\beq
 F^{(1)}_{\rm T3,A1}(x_i,\mu,\mu_f,Q^2)&=& \frac{\alpha_s(\mu_f) C_F}{8 \pi}
 \left [\frac{21}{2} \ln{\frac{\mu^2}{Q^2}} - 6 \ln{\frac{\mu^2_f}{Q^2}}
- \frac{53}{8} \ln{(x_1 x_2)} \right. \non
&&\left. - \frac{23}{8} \ln{x_1} - \frac{4}{9} \ln{x_2} - \frac{1}{4} \ln^{2}{x_2} - \frac{137}{48} \pi^{2}
+ \frac{337}{32} \right ], \label{eq:f1a1} \\
F^{(1)}_{\rm T3,A2}(x_i,\mu,\mu_f,Q^2)&=& \frac{\alpha_s(\mu_f) C_F}{8 \pi}
\left [\frac{21}{2} \ln{\frac{\mu^2}{Q^2}} - 6 \ln{\frac{\mu^2_f}{Q^2}}
- 8 \ln{(x_1 x_2)} \right.\non
&&\left. + \ln^{2}{x_{1}} + 4 \ln{x_2} - \frac{27}{12} \pi^{2} + \frac{1}{2} \ln{2} + 11 \right ].
\label{eq:f1a2}
\eeq

The IR-finite and $k_T$ dependent NLO hard kernel $H^{(1)}(\mu,\mu_f,x_i,Q^2) $
as given in Eq.~(\ref{eq:nlohk2}) describe the NLO twist-3 contribution to the LO
twist-3 hard kernel $H_a^{(0)}$ as given in Eq.~(6) for the Fig.1(a).
One can obtain the NLO twist-3 corrections to the LO twist-3 hard kernel $H_{b}^{(0)}$,
$H_{c}^{(0)}$ and $H_{d}^{(0)}$ for other three subdiagrams  Fig.1(b), 1(c) and 1(d)
respectively, by simple replacements.
For Fig.~1(b), for example, the two factors of the NLO twist-3 contributions
$F^{(1)}_{\rm T3,B1}(x_i,\mu,\mu_f,Q^2)$ and $F^{(1)}_{\rm T3,B2}(x_i,\mu,\mu_f,Q^2)$ can be
obtained from those in Eqs.~(\ref{eq:f1a1},\ref{eq:f1a2}) by simple replacements
$x_1 \longleftrightarrow x_2$:
\beq
F^{(1)}_{\rm T3,B1}(x_i,\mu,\mu_f,Q^2)&=& \frac{\alpha_s(\mu_f) C_F}{8 \pi}
 \left [\frac{21}{2} \ln{\frac{\mu^2}{Q^2}} - 6 \ln{\frac{\mu^2_f}{Q^2}}
- \frac{53}{8} \ln{(x_1 x_2)} \right. \non
&&\left. - \frac{23}{8} \ln{x_2} - \frac{4}{9} \ln{x_1} - \frac{1}{4} \ln^{2}{x_1} - \frac{137}{48} \pi^{2}
+ \frac{337}{32} \right ], \label{eq:f1b1} \\
F^{(1)}_{\rm T3,B2}(x_i,\mu,\mu_f,Q^2)&=& \frac{\alpha_s(\mu_f) C_F}{8 \pi}
\left [\frac{21}{2} \ln{\frac{\mu^2}{Q^2}} - 6 \ln{\frac{\mu^2_f}{Q^2}}
- 8 \ln{(x_1 x_2)} \right.\non
&&\left. + \ln^{2}{x_2} + 4 \ln{x_1} - \frac{27}{12} \pi^{2} + \frac{1}{2} \ln{2} + 11 \right ].
\label{eq:f1b2}
\eeq
We can also obtain the factors $F^{(1)}_{\rm T3,C1}$ and
$F^{(1)}_{\rm T3,C2}$ for subdiagrams  Fig.1(c)
by the replacements: $x_1 \rightarrow \bar{x}_1=1-x_{1}$ and
$x_2 \rightarrow \bar{x}_2=1-x_{2}$
from Eqs.~(\ref{eq:f1a1},\ref{eq:f1a2}).
For Fig.~1(d), finally, one finds the factors $F^{(1)}_{\rm T3,D1}$
and $F^{(1)}_{\rm T3,D2}$
from those in Eqs.~(\ref{eq:f1a1},\ref{eq:f1a2}) by the replacements:
$x_1 \rightarrow \bar{x}_2$ and $x_2 \rightarrow \bar{x}_1$.

\section{Numerical Analysis}

In this section, by employing the $\kt$ factorization theorem, we will calculate the
pion electromagnetic form factor $F^+(q^2)$ of the $\pi \gamma^* \to \pi$ process
numerically. Besides the LO twist-2 and twist-3 contributions, the NLO twist-2
contribution as given in Ref.~\cite{prd83-054029} and the NLO twist-3 contributions
evaluated in this paper are all taken into account.
We will compare the relative strength of the four parts numerically.

In order to compare our results directly with the theoretical predictions
for the LO twist-2, LO twist-3 and NLO twist-2 contributions to pion form
factor as presented in Ref.~\cite{prd83-054029}, we here firstly consider two different choices for
the pion distribution amplitudes (DA's): Set-A: the simple asymptotic pion DA's:
\beq
\phi_{\pi}^{A}(x) &= &\frac{3 f_{\pi}}{ \sqrt{6}} x (1-x),  \quad
\phi_{\pi}^{P}(x) = \frac{f_{\pi}}{2 \sqrt{6}}, \quad
\phi_{\pi}^{T}(x) =  \frac{f_{\pi}}{2 \sqrt{6}}(1-2x);
\label{eq:phipi01}
\eeq
with the pion decay constant $f_\pi=0.13$ GeV;  and Set-B: the nonasymptotic pion DA's
the same as those given in Eq.~(39) of Ref.~\cite{prd83-054029}:
\beq
\phi_{\pi}^{A}(x) &= &\frac{3 f_{\pi}}{ \sqrt{6}} x (1-x)
\left [ 1 +  0.16 C_2^{\frac{3}{2}}(u) + 0.04  C_4^{\frac{3}{2}}(u)\right ],
\non
\phi_{\pi}^{P}(x) &= &\frac{f_{\pi}}{2 \sqrt{6}} \left[ 1 + 0.59 C_2^{\frac{1}{2}}(u)
+ 0.09\; C_4^{\frac{1}{2}}(u) \right ], \non
\phi_{\pi}^{T}(x) &= & \frac{f_{\pi}}{2 \sqrt{6}} (1-2x)
\left [1 + 0.019 \left (1-10 x + 10 x^2 \right )  \right ], \label{eq:phipi02}
\eeq
where $u = 1-2 x $,  the Gegenbauer polynomials $C_{2,4}^{1/2,3/2}(u)$
can be found easily in Refs.~\cite{prd71-014015,jhep0605-004}.

In order to check the variations of the theoretical predictions induced by using different nonasymptotic
pion DA's, we also consider the third choice of pion DA's: Set-C: the pion DA's popularly used in
recent years, for example, in Refs.\cite{li2005,xiao2008,xiao2013}:
\beq
\phi_{\pi}^{A}(x) &= &\frac{3 f_{\pi}}{ \sqrt{6}} x (1-x)
\left [ 1 +  a_2^{\pi} C_2^{\frac{3}{2}}(u) + a_4^{\pi} C_4^{\frac{3}{2}}(u)
                    \right ], \non
\phi_{\pi}^{P}(x) &= &\frac{f_{\pi}}{2 \sqrt{6}} \left[ 1 + \left (30 \eta
_3 - \frac{5}{2} \rho_{\pi}^2 \right ) C_2^{\frac{1}{2}}(u) - 3 \left (\eta_3 \omega_3
+ \frac{9}{20} \rho_{\pi}^2 \left (1 + 6 a_2^{\pi}\right ) \right ) C_4^{\frac{1}{2}}(u) \right ],  \non
\phi_{\pi}^{T}(x) &= & \frac{f_{\pi}}{2 \sqrt{6}} (1-2x)
\left [1 + 6 \left (5 \eta_3 - \frac{1}{2} \eta_3 \omega_3 - \frac{7}{20} \rho_{\pi}^2
- \frac{3}{5} \rho_{\pi}^2 a_2^{\pi} \right ) \left (1-10 x + 10 x^2 \right )  \right ],
\label{eq:phipi03}
\eeq
where the Gegenbauer moments $a_i^\pi$, the parameters $\eta_3, \omega_3$ and $\rho_\pi$ are adopted from
Refs.~\cite{prd71-014015,jhep0605-004,li2005}:
\beq
a_2^{\pi} &=& 0.25, \quad a_4^{\pi} = -0.015, \quad
\rho_{\pi} = m_{\pi}/m_0, \quad \eta_3 = 0.015, \quad \omega_3 = -3.0,
\label{eq:input1}
\eeq
with $m_\pi=0.13$ GeV,  $m_0=1.74$ GeV.
It is easy to see that the asymptotic pion DA's in Eq.~(\ref{eq:phipi01}) are just the first (leading)
term of the nonasymptotic pion DA's as given in Eq.~(\ref{eq:phipi02}) and Eq.~(\ref{eq:phipi03}).
We will make numerical calculations by employing these three sets of pion DA's respectively,
for the sake of comparison and for the examination of the effects of the shape of the pion DA's.

When both the LO twist-2 and LO twist-3 contributions are included, the LO form factor
for  $\pi \gamma^{\star} \to \pi$ process can be written as
~\cite{prd83-054029,prd67-094013},
\beq
F^{+}(Q^2)|_{\rm LO} &=& \frac{8}{9} \pi Q^2 \int{dx_1 dx_2} \int{b_1 db_1 b_2 db_2}\non
&&\cdot \Bigl \{ x_{1} \phi^{A}_{\pi}(x_1) \phi^{A}_{\pi}(x_2)
 -  2 r_\pi^2\; \phi^{P}_{\pi}(x_2)\left [ (x_1 - 1) \phi^{P}_{\pi}(x_1)
+ (x_1 + 1) \phi^{T}_{\pi}(x_1) \right] \Bigr\} \non
&& \cdot \alpha_s(t) \cdot e^{-2 S_{\pi}(t)}\cdot S_t(x_2)\cdot h(x_1,x_2,b_1,b_2),
\label{eq:ff1}
\eeq
where $r_\pi^2=m_0^2/Q^2$, the first term  $x_1 \phi^{A}_{\pi}(x_1) \phi^{A}_{\pi}(x_2)$
leads to the LO twist-2 contribution, while the second term  in the large bracket
provide the LO twist-3 part. The $k_{T}$ resummation factor $S_{\pi}(t)$ is
adopted from Ref.~\cite{prd57-443,prd67-094013}
\beq
S_{\pi}(\mu, b_{i})= s(x_{i} \frac{Q}{\sqrt{2}},b_{i}) + s(\bar{x}_{i} \frac{Q}{\sqrt{2}}, b_{i})
                   + 2 \int^{\mu}_{1/b_{i}} \frac{d \bar{\mu}}{\bar{\mu}} r_Q(g(\bar{\mu})),
\eeq
with $i=1,2$ for the initial and final $\pi$ meson respectively.
The expressions of the function $s(Q',b)$ and the anomalous dimension
$\gamma_{q}$ can be found in Ref.~\cite{prd57-443}.
The threshold resummation factor $S_t(x)$ in Eq.~(\ref{eq:ff1}) is adopted
from Ref.~\cite{prd65-014007}
\beq
S_t(x)=\frac{2^{1+2c}\Gamma(3/2+c)}{\sqrt{\pi}\Gamma(1+c)}[x(1-x)]^c,
\eeq
and we here set the  parameter $c=0.3$.
The hard function $h(x_1,x_2,b_1,b_2)$ in Eq.~(\ref{eq:ff1})
comes form the Fourier transformation and can be written as \cite{prd83-054029}
\beq
h(x_1,x_2,b_1,b_2)&=&K_0\left (\sqrt{x_1x_2} Q b_1 \right ) \Bigl [\theta(b_1-b_2)I_0\left (\sqrt{x_2} Q b_2\right )
K_0\left (\sqrt{x_2} Q b_1 \right)\non
&& +\theta(b_2-b_1)I_0\left (\sqrt{x_2} Q b_1 \right )
K_0\left (\sqrt{x_2 } Q b_2\right )\Bigr],
\eeq
where $J_0$ is the Bessel function, and $K_0$, $K_1$ and $I_0$ are modified
Bessel functions.

According to the discussions as presented in Ref.~\cite{prd83-054029}, we get to
know that the relative strength of the NLO twist-2 contribution to the LO
twist-2 one has a moderate dependence on the choice of the renormalization scale $\mu$,
the factorization scale $\mu_f$ and the hard scale $t$.
One can see from the curves in the Fig.~6 of Re.~\cite{prd83-054029} that when one adopt
the conventional choice of the scales \cite{prd63-074009}, i.e.,
\beq
\mu=\mu_f= t = \max\left (\sqrt{x_1} Q, \sqrt{x_2}Q, 1/b_1,1/b_2 \right),
\label{eq:mumuf}
\eeq
where the hard scale $t$ is the largest energy scale in Fig.~\ref{fig:fig1},
the NLO twist-2 correction becomes less than $40\%$ of the LO twist-2 contribution
as $Q^2 > 7$ GeV$^2$, or less than $20\%$ of the total LO contribution.
This means that such choice can minimize  the NLO twist-2 correction to
the form factors in consideration. We here also make the same choices as given in
Eq.~(\ref{eq:mumuf}) in our numerical  calculations of the NLO twist-2 and twist-3
contributions. For more details about the choice of $\mu$, $\mu_f$ and hard scale $t$,
one can see Ref.~\cite{prd83-054029}.

When the LO twist-2, LO twist-3, NLO twist-2 and NLO twist-3 contributions to the
pion form factors are all taken into account, the pion form factor $F^+(q^2)$ for
$\pi \gamma^{\star} \to \pi$  process in the $k_T$ factorization can be written as
\beq
F^{+}(Q^2)|_{\rm NLO} &=& \frac{8}{9} \pi Q^2 \int{dx_1 dx_2} \int{b_1 db_1 b_2 db_2}
\cdot \Bigl \{ x_{1}\; \phi^{A}_{\pi}(x_1) \phi^{A}_{\pi}(x_2)
\cdot \left [ 1+ F^{(1)}_{\rm T2}(x_i,t,Q^2)\right ] \non
&&  - 2 r_\pi^2\; x_1 \; \phi^{P}_{\pi}(x_2)
  \left [ 1+  \cdot F^{(1)}_{\rm T3}(x_i,t,Q^2)\right]
  \left (\phi^{P}_{\pi}(x_1) + \phi^{T}_{\pi}(x_1) \right ) \non
&& + 2 r_\pi^2\;  \phi^{P}_{\pi}(x_2)
  \left[ 1+ \ov{F}^{(1)}_{\rm T3}(x_i,t,Q^2)\right ]
  \cdot \left (\phi^{P}_{\pi}(x_1) - \phi^{T}_{\pi}(x_1) \right) \Bigr\} \non
&&\cdot \alpha_s(t) \cdot e^{-2 S_{\pi}(t)}\cdot S_t(x_2)\cdot h(x_1,x_2,b_1,b_2)£¬
\label{eq:ff2}
\eeq
where the factor $F^{(1)}_{\rm T2}(x_i,t,Q^2)$ denotes the NLO twist-2 contribution
as given in Ref.~\cite{prd83-054029}
\beq
F^{(1)}_{\rm T2}(x_i,t,Q^2)&=& \frac{\alpha_s(t) C_F}{4 \pi}
\left [- \frac{3}{4} \ln{\frac{t^2}{Q^2}}
- \ln^{2}{x_{1}} - \ln^{2}{x_{2}} + \frac{45}{8}\ln{x_{1}}\ln{x_{2}} \right. \non
&&\left. + \frac{5}{4} \ln{x_1} + \frac{77}{16} \ln{x_2} + \frac{1}{2} \ln{2} + \frac{5}{48} \pi^{2}
+ \frac{53}{4} \right ].
\label{eq:ffnlot2}
\eeq
The factor $F^{(1)}_{\rm T3}(x_i,t,Q^2)$ and $\ov{F}^{(1)}_{\rm T3}(x_i,t,Q^2)$
in Eq.~(\ref{eq:ff2}) describe the NLO twist-3 contributions and have been defined in
Eqs.~(\ref{eq:f1a1},\ref{eq:f1a2}). By making the same choice of scales
$(\mu,\mu_f,t)$ as the one in Eq.~(\ref{eq:mumuf}), these two factors become relatively
simple
\beq
F^{(1)}_{\rm T3}(x_i,t,Q^2)&=& \frac{\alpha_s(t) C_F}{4 \pi}
 \Bigl [\frac{9}{4} \ln{\frac{t^2}{Q^2}} - \frac{53}{16} \ln{x_1x_2}
 - \frac{23}{16} \ln{x_1} \non
&&  - \frac{2}{9} \ln{x_2} - \frac{1}{8} \ln^{2}{x_2} -
\frac{137}{96} \pi^{2} + \frac{337}{64} \Bigr ], \label{eq:f1a1b} \\
\ov{F}^{(1)}_{\rm T3}(x_i,t,Q^2)&=& \frac{\alpha_s(t) C_F}{4 \pi}
\Bigl [\frac{9}{4} \ln{\frac{t^2}{Q^2}} - 4 \ln{(x_1 x_2)}
+ \frac{1}{2}\ln^{2}{x_{1}} \non
&& + 2 \ln{x_2} - \frac{27}{24} \pi^{2} + \frac{1}{4} \ln{2} + \frac{11}{2} \Bigr ].
\label{eq:f1a2b}
\eeq

%%--------------------------------------------------------------
\begin{table}[tb]
\begin{center}
\caption{ The theoretical predictions for contributions to
$Q^2F^+(Q^2)$ from different orders and twists, for $Q^2=(1,3,5,7,10,100)$
GeV$^2$ and for different cases: i.e., using different sets of pion DA's respectively.}
\label{tab:001}
\vspace{0.2cm}
\begin{tabular}{l |c|c c c cc| c} \hline \hline
$Q^2F^+(Q^2)$ & DA's& 1    & 3    & 5    & 7    & 10     &100 \\ \hline
LO T-2 &Set-A&0.078 & 0.070&0.080 &0.085 &0.089  &0.086 \\
       &Set-B&0.075 & 0.075&0.086 &0.093 &0.098  &0.109 \\
       &Set-C&0.071 & 0.076&0.089 &0.095 &0.102  &0.116 \\ \hline
NLO-T2 &Set-A&0.102 & 0.039&0.035 &0.034 &0.033  &0.022 \\
       &Set-B&0.106 & 0.044&0.040 &0.039 &0.037  &0.030 \\
       &Set-C&0.106 & 0.047&0.043 &0.041 &0.040  &0.033 \\ \hline
LO-T3  &Set-A&0.399 & 0.141&0.099 &0.076 &0.056  &0.007 \\
       &Set-B&0.905 & 0.558&0.491 &0.447 &0.405  &0.207 \\
       &Set-C&0.519 & 0.206&0.145 &0.113 &0.087  &0.010 \\ \hline
NLO-T3 &Set-A &-0.252&-0.059&-0.037&-0.028&-0.021&-0.003 \\
       &Set-B &-0.261&-0.155&-0.151&-0.148&-0.144&-0.104 \\
       &Set-C &-0.286&-0.073&-0.046&-0.034&-0.025&-0.002 \\ \hline
Full LO&Set-A &0.476 & 0.211&0.179 &0.160 &0.145 &0.092 \\
       &Set-B &0.981 & 0.633&0.571 &0.540 &0.504 &0.316 \\
       &Set-C &0.590 & 0.282&0.234 &0.209 &0.189 &0.126 \\ \hline
LO+NLO &Set-A &0.326 & 0.190&0.176 &0.166 &0.156 &0.111 \\
       &Set-B &0.825 & 0.522&0.466 &0.431 &0.397 &0.242 \\
       &Set-C &0.409 & 0.256&0.230 &0.216 &0.204 &0.157 \\ \hline \hline
\end{tabular}
\end{center} \end{table}

%%--------------------------------------------------------------
\begin{table}[tb]
\begin{center}
\caption{ The ratios of the different contributions or their combinations
as defined in Eq.~(\ref{eq:r1234}), for $Q^2=(1,3,5,7,10,100)$ GeV$^2$
respectively.}
\label{tab:002}
\vspace{0.2cm}
\begin{tabular}{l |c|c c c cc| c} \hline \hline
Ratios       & DA's & 1 & 3& 5 & 7& 10 &100 \\ \hline

$R_1$ &Set-A&1.303  & 0.549 &0.440  &0.398 &0.367   &0.256 \\
      &Set-B&1.406  & 0.595 &0.467  &0.418 &0.382   &0.273 \\
      &Set-C&1.488  & 0.616 &0.481  &0.430 &0.393   &0.284 \\ \hline
$R_2$ &Set-A&-0.633 &-0.423 &-0.376 &-0.368&-0.372  &-0.487 \\
      &Set-B&-0.288 &-0.277 &-0.307 &-0.331&-0.356  &-0.503 \\
      &Set-C&-0.552 &-0.356 &-0.319 &-0.302&-0.288  &-0.199 \\ \hline
$R_3$ &Set-A&-0.316 &-0.099 &-0.012 & 0.036& 0.080 &0.203 \\
      &Set-B&-0.158 &-0.174 &-0.191 &-0.202&-0.212 &-0.235 \\
      &Set-C&-0.306 &-0.094 &-0.016 & 0.032& 0.080 &0.247 \\\hline
$R_4$ &Set-A&-0.462 &-0.110 &-0.013 & 0.035& 0.074  &0.169 \\
      &Set-B&-0.188 &-0.211 &-0.237 &-0.254&-0.269  &-0.307 \\
      &Set-C&-0.441 &-0.103 &-0.016 & 0.031& 0.074  &0.198 \\ \hline \hline
\end{tabular}
\end{center} \end{table}

\begin{figure*}
\centering
\vspace{-0.5cm}
\includegraphics[width=0.46\textwidth]{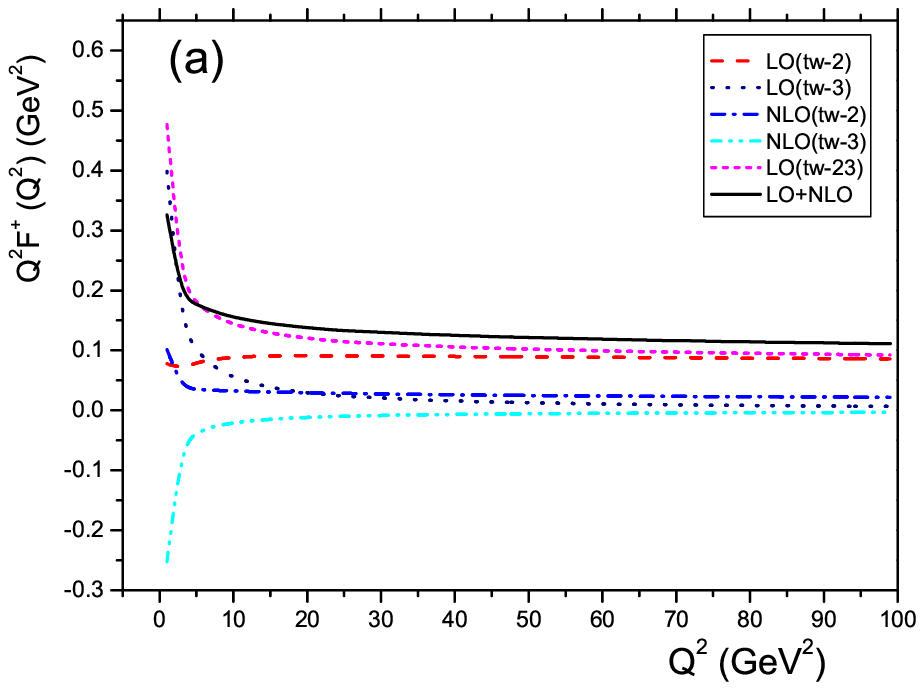}
\includegraphics[width=0.46\textwidth]{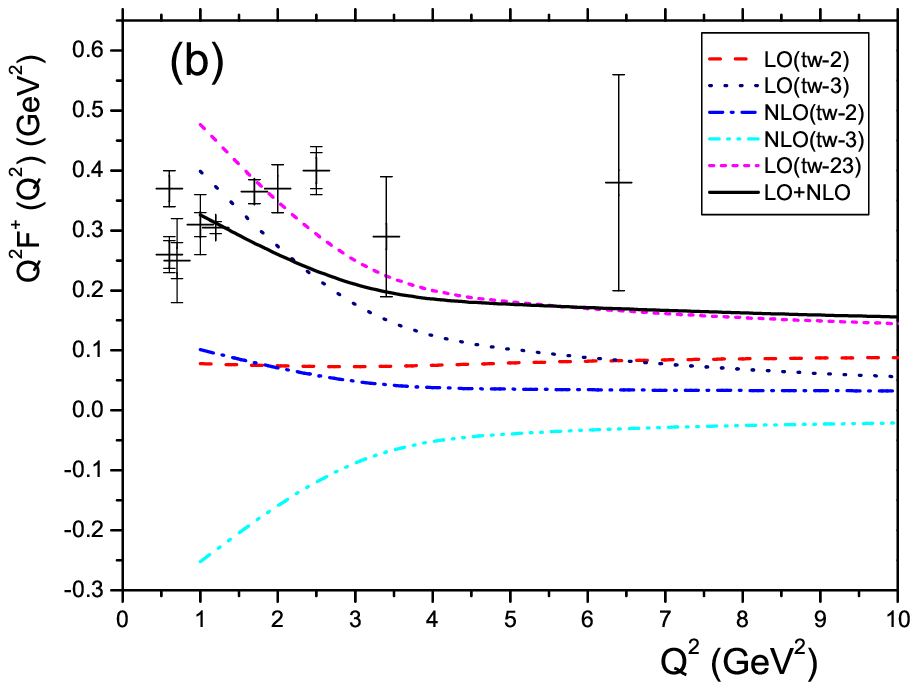}
\vspace{0cm}
\caption{Contributions to $Q^2 F^+(Q^2)$ from different orders and twists, using the asymptotic
pion DA's as given in Eq.~(\ref{eq:phipi01}).
Figure 7(a) shows the $Q^2$-dependence for $1\leq Q^2 \leq 100$ GeV$^2$,
while 7(b) is the enlargement of  7(a) in the low-$Q^2$ region: $1\leq Q^2\leq 10$ GeV$^2$.
The experiment data  in 7(b) are taken from Refs.~\cite{prd17-1693,prc78-045203}.}
\label{fig:fig7}
\end{figure*}

\begin{figure*}
\centering
\vspace{-0.5cm}
\includegraphics[width=0.46\textwidth]{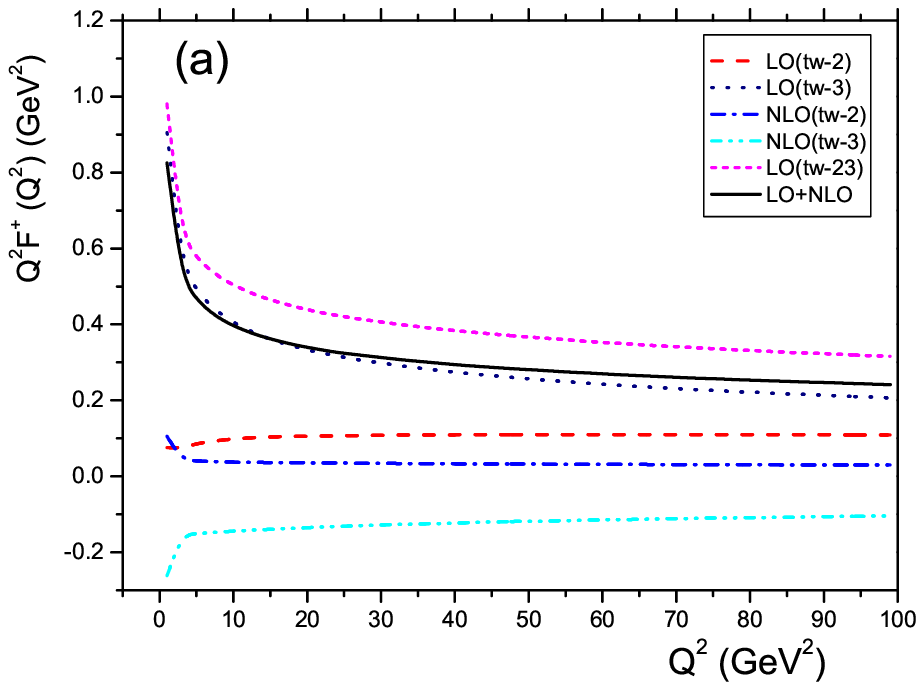}
\includegraphics[width=0.46\textwidth]{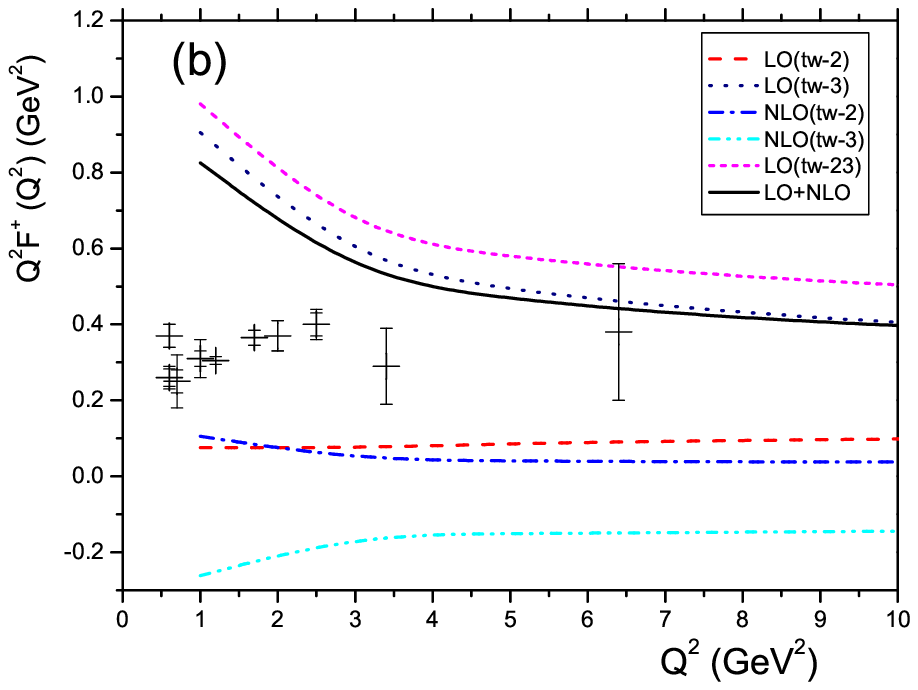}
\vspace{0cm}
\caption{The same as Fig.~7, but using the nonasymptotic pion DA's as given in  Eq.~(\ref{eq:phipi02}).}
\label{fig:fig8}
\end{figure*}

\begin{figure*}
\centering
\vspace{-0.5cm}
\includegraphics[width=0.46\textwidth]{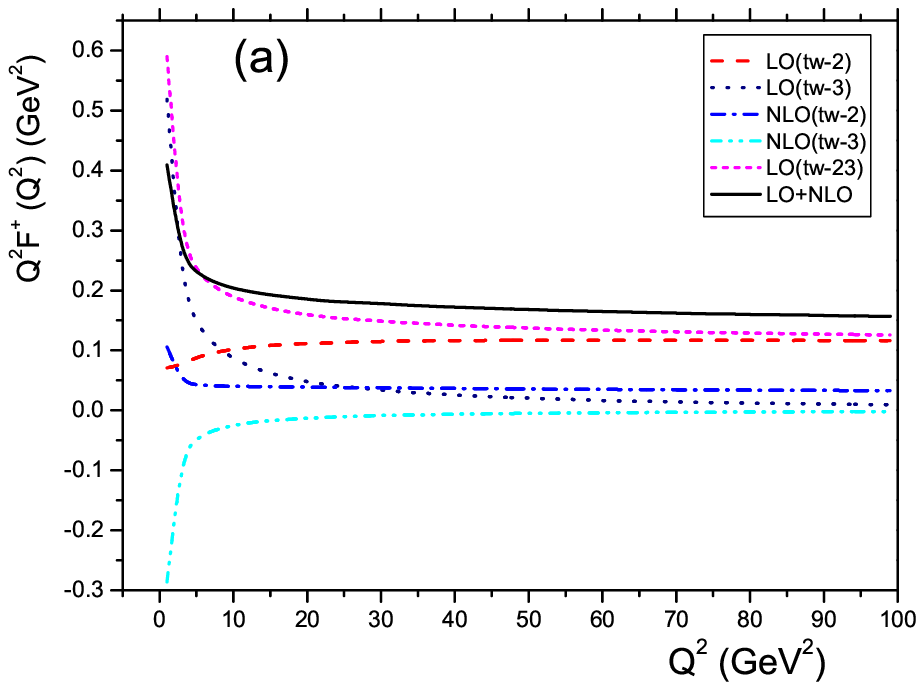}
\includegraphics[width=0.46\textwidth]{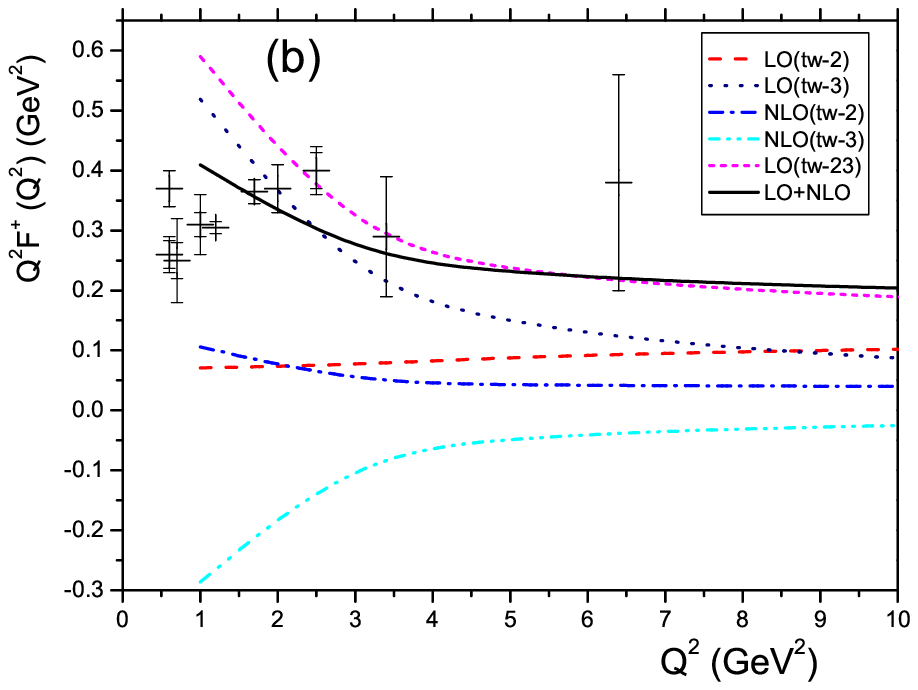}
\vspace{0cm}
\caption{The same as Fig.~7, but using the nonasymptotic pion DA's as given in Eq.~(\ref{eq:phipi03}).}
\label{fig:fig9}
\end{figure*}

\begin{figure*}
\centering
%%\vspace{-0.5cm}
\includegraphics[width=0.32\textwidth]{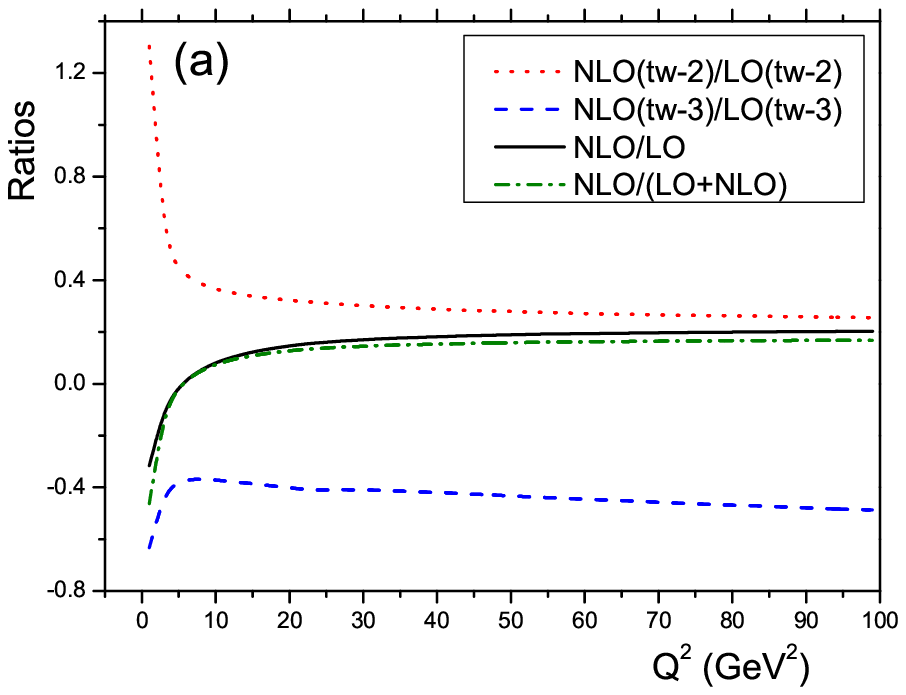}
\includegraphics[width=0.32\textwidth]{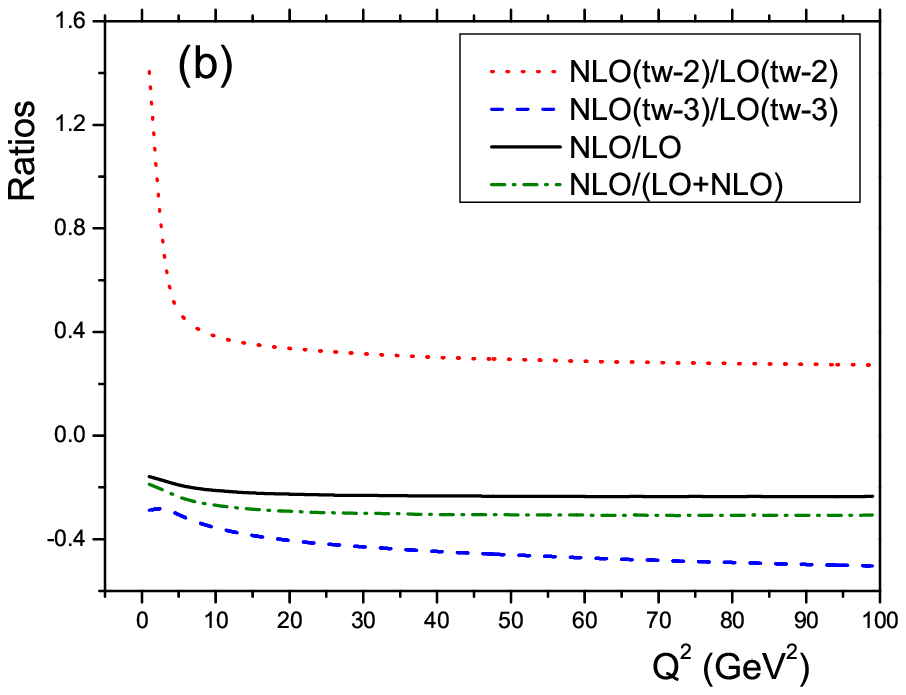}
\includegraphics[width=0.32\textwidth]{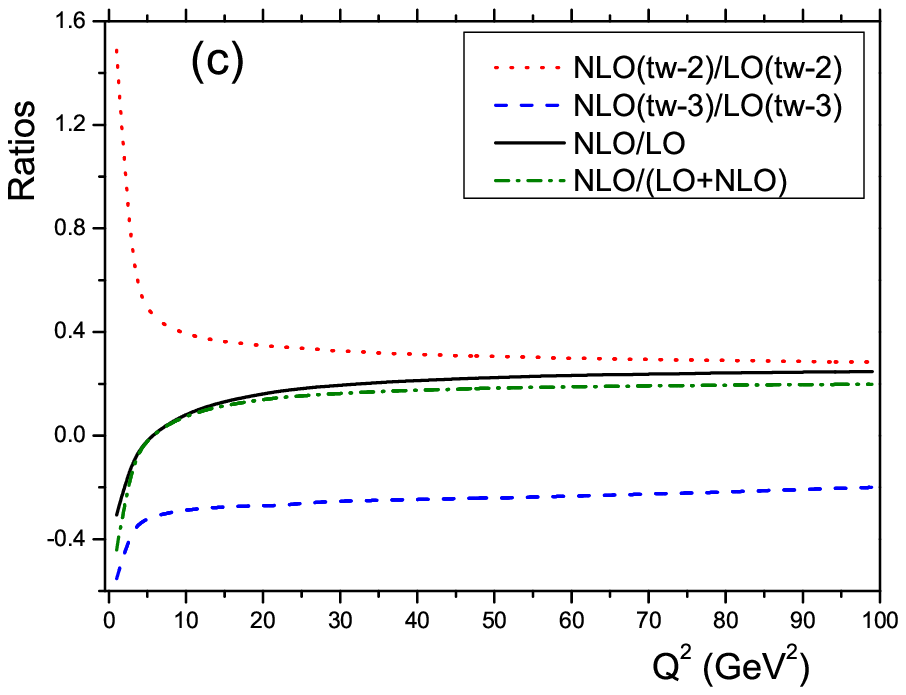}
%\vspace{0cm}
\caption{Ratios of the NLO corrections over the LO contributions to the
pion form factor, assuming $\mu = \mu_f = t$.
(a) the asymptotic pion DA's in Eq.~(\ref{eq:phipi01}) are used; (b) the nonasymptotic DA's in
Eq.(\ref{eq:phipi02}) are used; and (c) the nonasymptotic DA's in Eq.(\ref{eq:phipi03})  are used.}
\label{fig:fig10}
\end{figure*}

By using the three sets of pion distribution amplitudes $\phi_\pi^{A,P,T}(x)$
as given in Eqs.~(\ref{eq:phipi01}), (\ref{eq:phipi02}) and Eq.~(\ref{eq:phipi03})
respectively, and fixing the scales as in Eq.(\ref{eq:mumuf}),
we calculate the four different LO and NLO contributions to the pion form factors
and show the theoretical predictions in Table \ref{tab:001} and  \ref{tab:002},
and in Figs.~\ref{fig:fig7}-\ref{fig:fig10}, respectively.

In Table \ref{tab:001}, we list the theoretical predictions for the four kinds of
contributions: the LO twist-2, LO twist-3, NLO twist-2 and NLO twist-3 contributions
to the pion form factors $Q^2F^+(Q^2)$ for fixed values of $Q^2=(1,3,5,7,10,100)$
GeV$^2$. In the numerical calculations, three sets of different choices of pion DA's
are used respectively, with the labels of Set-A, Set-B and Set-C.

In order to compare the relative strength of different contributions directly
we define the following four ratios:
\beq
R_1&=&\frac{F^+_{\rm NLO-T2}(Q^2)}{F^+_{\rm LO-T2}(Q^2) },\quad
R_2=  \frac{F^+_{\rm NLO-T3}(Q^2)}{F^+_{\rm LO-T3}(Q^2) },\non
R_3&=&  \frac{F^+_{\rm NLO}(Q^2)}{F^+_{\rm LO}(Q^2)}, \quad
R_4=\frac{F^+_{\rm NLO}(Q^2)}{F^+_{\rm NLO}(Q^2)+F^+_{\rm LO}(Q^2)},
\label{eq:r1234}
\eeq
where $R_1$ ($R_2$) measures the ratio between NLO twist-2 (twist-3) and LO twist-2
(twist-3) contribution, $R_3$ describes the relative strength between the NLO contribution
and the LO ones, and finally $R_4$ is the ratio of the NLO contribution over
the total contribution: all four parts, LO plus NLO contributions.
In Table \ref{tab:002}, we present the numerical values of the ratios
of the different kind of contributions to $F^+(Q^2)$ for fixed values of
$Q^2=(1,3,5,7,10,100)$ GeV$^2$ and for three different sets of the pion DA's, respectively.

In Fig.~\ref{fig:fig7}, we show the $Q^2$-dependence of the various
contributions to the pion form factors from different orders and twists
for $1\leq Q^2 \leq 100$ GeV$^2$, by using the asymptotic pion DA's as given
in Eq.~(\ref{eq:phipi01}) and setting $\mu=\mu_f=t$.
The Fig.~7(b) shows the enlargement of Fig.~7(a) in the low-$Q^2$ region:
$1\leq Q^2\leq 10$ GeV$^2$. The experimental data shown in
Fig.~7(b) are taken from Refs.~\cite{prd17-1693,prc78-045203}.
The Fig.~\ref{fig:fig8} and Fig.~\ref{fig:fig9} also show the
$Q^2$-dependence of the various contributions to the pion form factors, but
using the nonasymptotic pion DA's as given in Eqs.~(\ref{eq:phipi02},\ref{eq:phipi03})
instead of the asymptotic ones in Eq.(\ref{eq:phipi01}).
In Fig.~\ref{fig:fig10}, we show the $Q^2$-dependence of the four ratios
$R_{1,2}$ and $R_{3,4}$ for $1\leq Q^2 \leq 100$ GeV$^2$, assuming $c=0.3$ and $\mu=\mu_f=t$
and employing the three different sets of the pion DA's.

%-----------------------------------------------------------------------
%%------------------------------------------------------------------------

%-----------------------------------------------------------------------
%%------------------------------------------------------------------------

%-----------------------------------------------------------------------
%%------------------------------------------------------------------------

%%-----------------------------------------------------------------------
%%------------------------------------------------------------------------

From the theoretical predictions for the pion form factors from different orders and twists,
as listed in Table \ref{tab:001} and \ref{tab:002},
and illustrated in Fig.~\ref{fig:fig7}-\ref{fig:fig10}, one can have the following observations:
\begin{enumerate}
\item[(i)]
For the LO twist-2 and NLO twist-2 contributions to the pion form factors $F^+(Q^2)$ obtained in this work
agree very well with those presented in Ref.~\cite{prd83-054029} when the same Set-A and Set-B pion DA's
are used, as can be seen easily from the numerical results in Table I and II, as well as in the
Figs.~7-9. Even when the Set-C pion DA's as given in Eq.~(\ref{eq:phipi03}) were used, the theoretical
predictions for the LO twist-2 and NLO twist-2 contributions are still well consistent with those
in Ref.~\cite{prd83-054029}, since the twist-2 $\phi_\pi^A(x)$ in Eq.~(\ref{eq:phipi02}) and
Eq.~(\ref{eq:phipi03}) have a little difference only. By using $a_2^\pi=0.25$ and $a_4^\pi=-0.015$, we find from
Eq.~(\ref{eq:phipi03}) directly that
\beq
\phi_\pi^A(x)=\frac{f_{\pi}}{2 \sqrt{6}} \left[ 1 + 0.25 C_2^{\frac{3}{2}}(u)
-0.015\; C_4^{\frac{3}{2}}(u) \right ].
\eeq
The coefficient of the second term is $a_2^\pi=0.25$, close to the $0.16$ in the $\phi_\pi^A(x)$ in Eq.~(\ref{eq:phipi02}).

\item[(ii)]
For the LO twist-3 and NLO twist-3 contributions, one can see from the numerical results in Table I
and the cures in Figs.7-9 that these two contributions are rather similar with each other in both the magnitude and the shape
when Set-A and Set-C pion DA's are used, respectively.
When the Set-B pion DA's as given in Eq.~(\ref{eq:phipi02}) are employed, however, the corresponding theoretical
predictions for both the LO twist-3 and NLO twist-3 contributions
become rather different from those obtained by using the Set-C pion DA's. The reason is that
there is a clear difference for the twist-3 DA's $\phi_\pi^P(x)$ and $\phi_\pi^T(x)$ in
Eq.~(\ref{eq:phipi02}) and Eq.~(\ref{eq:phipi03}), specifically for $\phi_\pi^T(x)$.
Using the Gegenbauer moments and other input parameters as given in Eq.~(\ref{eq:input1}), we find numerically that
\beq
\phi_{\pi}^{P}(x) &= &\frac{f_{\pi}}{2 \sqrt{6}} \left[ 1 + 0.43 C_2^{\frac{1}{2}}(u)
+ 0.11\; C_4^{\frac{1}{2}}(u) \right ], \non
\phi_{\pi}^{T}(x) &= & \frac{f_{\pi}}{2 \sqrt{6}} (1-2x) \left [1 + 0.56 \left (1-10 x + 10 x^2 \right )  \right ].
\label{eq:phipi03b}
\eeq
One can see that the coefficients $(0.43,0.11)$ of $\phi_\pi^P(x)$ in
Eq.(\ref{eq:phipi03b}) are close to $(0.59,0.09)$ in Eq.(\ref{eq:phipi02}),
but  the coefficient $0.56$ of $\phi_\pi^T(x)$ in Eq.(\ref{eq:phipi03b}) is much larger
than  $0.019$ in Eq.~(\ref{eq:phipi02}). Because the coefficient $0.019$ is too small,
the twist-3
nonasymptotic $\phi_\pi^T(x)$ as given in Eq.~(\ref{eq:phipi02}) is in
fact the same one as the asymptotic
$\phi_\pi^T(x)$ as given in Eq.~(\ref{eq:phipi01}). This is a little
unreasonable in our opinion.

\item[(iii)]
For the LO twist-2 contribution $F^+_{\rm LO-T2}(Q^2)$, the theoretical prediction remain stable in  the whole
range of $1\leq Q^2\leq 100$ GeV$^2$
when asymptotic $\phi^A(x)$ is used. While it becomes a little bit large along with the increase of $Q^2$
when  other two sets pion DA's are employed, since the twist-2 $\phi_\pi^A(x)$
in Eqs.(\ref{eq:phipi02},\ref{eq:phipi03}) are very similar with each other.

\item[(iv)]
For the NLO twist-2 contribution, the value of   $Q^2 F^+_{\rm NLO-T2}(Q^2)$ becomes smaller rapidly in the
low-$Q^2$ region, say $1\leq Q^2 \leq 3$ GeV$^2$,
and than decrease slowly from $\sim 0.044$ to $0.030$ along with the increase of $Q^2$ from $3$ to $100$ GeV$^2$.
The ratio $R_1$ is changing from $\sim 60\%$ for $Q^2=3$ GeV$^2$ to $\sim 26\%$ for $Q^2=100$ GeV$^2$.

\item[(v)]
For the LO twist-3 contribution, the theoretical predictions for  $Q^2 F^+_{\rm LO-T3}(Q^2)$
obtained by using the Set-C pion DA's are about $15\%$  larger than
those obtained when the asymptotic $\phi_\pi^{P,T}$ are used, but much smaller than the ones from
the Set-B pion DA's. The reason is of course the special choice of $\phi_\pi^T(x)$ in Set-B pion DA's.
In the low-$Q^2$ region of $Q^2 < 10$ GeV$^2$, the LO twist-3 contribution becomes small rapidly.
From the numbers in Table I for the case of Set-C pion DA's, one can see that the ratio between the LO twist-3 and
LO twist-2 contribution are approximately $7.3,1.6,0.9$ for $Q^2=1,5,10$ GeV$^2$ respectively.
This is rather different from the behavior when Set-B pion DA's are used, in which
the LO twist-3 part is always larger than the LO twist-2 contribution by a factor  $\geq 4.1$.

\item[(vi)]
For the NLO twist-3 contribution, the theoretical predictions for  $Q^2 F^+_{\rm NLO-T3}(Q^2)$
has an opposite sign with its counterpart $Q^2 F^+_{\rm NLO-T2}(Q^2)$ and largely canceled each other.
The NLO twist-3 contribution calculated by using the Set-A and Set-C pion DA's are similar in size ( the difference
is around $10\%$ ) in the whole range of $Q^2$ and become smaller rapidly along with the increase of $Q^2$,
as illustrated by the lowest dot-dash curves in Figs.\ref{fig:fig7} and \ref{fig:fig9}.
The $Q^2 F^+_{\rm NLO-T3}(Q^2)$ from  the Set-B pion DA's  is similar in size with those for other two cases
at the starting point $Q^2=1$ GeV$^2$, but remain basically stable in the range of $Q^2 > 3$ GeV$^2$.

\item[(vii)]
The ratio $R_1$ from the three different sets of pion DA's has similar value
and $Q^2$-dependence, as illustrated by the upper dots curves in Fig.~\ref{fig:fig10}.
The other three ratios $R_{2,3,4}$ as shown in Fig.~\ref{fig:fig10}(a)
and  \ref{fig:fig10}(c) are also similar in size and in their $Q^2$-dependence,
but rather different from those obtained by
using the Set-B pion DA's. The ratio $R_2$ in Fig.~\ref{fig:fig10}(c), for example,
is changing from $-0.552$ for $Q^2=1$ GeV$^2$ to $-0.199$ for $Q^2=100$ GeV$^2$,
while the ratio $R_2$ in Fig.~\ref{fig:fig10}(b) changes its value from
$-0.288$ for $Q^2=1$ GeV$^2$ to $-0.503$ for $Q^2=100$ GeV$^2$.

\item[(viii)]
When the Set-C pion DA's are used, one can see from the Table \ref{tab:002} and Fig.~\ref{fig:fig10}(c) that
(a) at twist-2 level, the NLO twist-2 contribution can provide a strong enhancement to the LO twist-2 part,
from $30\%$ to $60\%$ in the range of $3< Q^2 \leq 100 $ GeV$^2$;
(b) at twist-3 level, the NLO twist-3 contribution is about $30\%$ of the LO  twist-3 part in magnitude
in the range of $3 < Q^2 \leq 10$ GeV$^2$, but has an opposite sign with its LO counterpart in the whole
range of $Q^2$, which leads to a partial cancelation of the LO and NLO twist-3 contributions;

\item[(ix)]
Because of the strong cancelation between the NLO twist-2 and NLO twist-3 contributions,
the total NLO contribution to pion form factor $F^+(Q^2)$ become small in size
with respect to the total LO part,
from about $-31\%$  for $Q^2=1$ GeV$^2$ to $\sim 25\%$ for $Q^2=100$ GeV$^2$ when
the Set-C pion DA's are used.
The ratio $R_3$ change its sign at the point $Q^2\sim 6$ GeV$^2$,
as shown by the solid curve in Fig.~\ref{fig:fig10}(c).
When the Set-B pion DA's are used, however, the ratio $R_3$ is always negative
and keep stable in size for the whole range of $Q^2$.

\end{enumerate}

\section{Conclusion}

In this paper, we made the first calculation for the NLO twist-3 contributions to the pion
electromagnetic form factor for the $\pi \gamma^* \to \pi$  process, by employing the $\kt$
factorization theorem and using the nonasymptotic pion distribution amplitudes: the leading twist-2
$\phi_\pi^A(x)$ the twist-3 $\phi_\pi^{P,T}(x)$.

The UV divergences at the NLO twist-3 level are found to be the same ones as the NLO twist-2 part,
which confirms the universality of the non-perturbative wave functions.
These UV divergences are renomalized into the coupling constants and quark fields.
Both the soft and collinear divergences in the NLO QCD quark diagrams and in the
NLO effective diagrams for pion wave functions are regulated by the off-shell
momentum $k^2_T$ of the light quark.
The soft divergences cancels themselves in the quark diagrams and the collinear
divergences cancels between the QCD quark diagrams and the effective diagrams
at twist-3, in cooperation with the cancelation at the leading twist-2
\cite{prd83-054029}, verified the validity of the $\kt$ factorization for
the exclusive decays at the NLO level.
The large double logarithm $\ln^2{x_i}$ in the NLO hard kernel are strongly suppressed
by the Sudakov factor, then the NLO corrections are under control.

From the analytical calculations  we obtained two
factors $F^{(1)}_{\rm T3}(x_i,t,Q^2)$ and $\ov{F}^{(1)}_{\rm T3}(x_i,t,Q^2)$, which describe
directly the NLO twist-3 contributions to the pion form factors $F^+(Q^2)$ as shown
in Eq.~(\ref{eq:ff2}).
From the numerical results and phenomenological analysis we found the following points:
\begin{enumerate}
\item[(i)]
For the LO twist-2, twist-3 and NLO twist-2 contributions, our results agree very well with those
as given in previous work \cite{prd83-054029} for both the magnitude and the $Q^2$-dependence of
the individual part.

\item[(ii)]
The newly calculated NLO twist-3 contribution is negative in sign and will interfere destructively
with the NLO twist-2 part, leaves a relatively small total NLO contribution, which can result in
a roughly $\pm 20\%$ corrections to the total LO contribution in almost all considered ranges of  $Q^2$.

\item[(iii)]
The theoretical predictions for $Q^2F^+(Q^2)$ in the low-$Q^2$ region agree well with currently available
data. The inclusion of NLO contributions results in a better agreement between the theory and the experiments.

\item[(x)]
The theoretical predictions for the pion form factors obtained by employing the $\kt$ factorization theorem
have a moderator dependence on the form and the shape of the pion distribution amplitudes, this is the main
source of the theoretical uncertainty.

\end{enumerate}

\section{Acknowledement}

The authors would like to thank Hsiang-nan Li, Cai-Dian Lu, Xin Yu, 
Yu-Ming Wang and Yue-Long Shen for collaborations and valuable discussions.
This work is supported by the National Natural Science Foundation
of China under Grant No.10975074,11235005,
and by the Project on Graduate Students Education and Innovation of Jiangsu Province
under Grant No. CXZZ13-0391.

%%---------------------------------------------------------------------------------------


\begin{thebibliography}{99}

\bibitem{npb325-62}
J.~Botts and G.~Sterman, \npb{\bf 325}, 62 (1989);
% Hard Elastic Scattering in QCD: Leading Behavior %
H.N.~Li and G.~Sterman, \npb {\bf 381}, 129 (1992);
%%  The perturbative pion form factor with Sudakov suppression.
T.~Huang and Q.X. Shen, \zpc {\bf 50}, 139 (1991);
%% The applicability of perturbative QCD to the pion form factor and the pionic wavefunction.
F.G.~Cao, T. Huang and C.W.~Luo, \prd {\bf 53}, 5358(1995).
%%Reexamination of the perturbative pion form factor with Sudakov suppression

\bibitem{npb360-3}
J.C.~Collins and R.K.~Ellis, \npb{\bf 360}, 3 (1991).
% Heavy quark production in very high-energy hadron collisions%

\bibitem{prl74-4388}
H.N.~Li and H.L.~Yu, \prl {\bf 74}, 4388 (1995);
\plb {\bf 353}, 301 (1995);
\prd {\bf 53}, 2480 (1996).
%Extraction of V(ub) from decay B ---> pi lepton neutrino%
%PQCD analysis of exclusive charmless B meson decay spectra%
%Perturbative QCD analysis of B meson decays%

\bibitem{plb504-6}
Y.Y.~Keum, H.N.~Li and A.I,Sanda, \plb {\bf 504}, 6 (2001),
\prd {\bf 63}, 054008 (2001).
%Fat penguins and imaginary penguins in perturbative QCD%
%Penguin enhancement and B ---> K pi decays in perturbative QCD%

\bibitem{prd63-074009}
C.D.~L\"{u},  K.~Ukai and M.Z.~Yang, \prd {\bf 63}, 074009 (2001).
%Branching ratio and CP violation of B ---> pi pi decays in perturbative QCD approach%

\bibitem{plb242-97}
S.~Catani, M.~Ciafaloni and F.~Hautmann, \plb{\bf 242}, 97 (1990), \npb{\bf 366}, 135 (1991).
%GLUON CONTRIBUTIONS TO SMALL x HEAVY FLAVOR PRODUCTION%
%High-energy factorization and small x heavy flavor production %

\bibitem{prl65-2343}
J.P.~Ralston and B.~Pire, \prl{\bf 65}, 2343 (1990).
%Quantum chromotransparency%

\bibitem{prd76-034008}
S.Nandi and H.N.~Li, \prd{\bf 76}, 034008 (2007).
%Next-to-leading-order corrections to exclusive processes in k(T) factorization%

\bibitem{prd83-054029}
H.N.~Li, Y.L.~Shen, Y.M.~Wang and H.Zou, \prd{\bf 83}, 054029(2011).
%Next-to-leading-order correction to pion form factor in k_T factorization%

\bibitem{prd85-074004}
H.N.~Li, Y.L.~Shen, Y.M.~Wang, \prd {\bf 85}, 074004 (2012).
%Next-to-leading-order corrections to B¡ú¦Ð form factors in kT factorization%

\bibitem{prd64-014019}
H.N.~Li, \prd{\bf 64}, 014019(2001).
%Perturbative QCD factorization of pi gamma* ---> gamma (pi) and B ---> gamma (pi) lepton anti-neutrino%

\bibitem{prd65-014007}
T.~Kurimoto, H.N.~Li, and A.I.~Sanda, \prd {\bf 65}, 014007 (2001).

\bibitem{epjc23-275}
C.D,~L\"{u} and M.Z.~Yang, \epjc{\bf23}, 275-287 (2002).

\bibitem{prd66-094010}
H.N.~Li, \prd {\bf 66}, 094010 (2002).
%Threshold resummation for exclusive B meson decays %

\bibitem{plb555-197}
H.N.~Li, \plb {\bf 555}, 197 (2003).
%Threshold resummation for nonleptonic B meson decays%

\bibitem{prd65-094032}
B.W.Harris and J.F.Owens, \prd {\bf 65}, 094032 (2002).
% The Two cutoff phase slicing method%

\bibitem{zpc48-239}
V.M.~Braun and I.E.~Filyanov, \zpc {\bf 48}, 239 (1990).
%Conformal Invariance and Pion Wave Functions of Nonleading Twist%

\bibitem{jhep9901-010}
P.~Ball, \jhep {\bf 9901}, 010 (1999).
%Theoretical update of pseudoscalar meson distribution amplitudes of higher twist: The Nonsinglet case%

\bibitem{jhep0605-004}
P.~Ball, V.M.Braun and A.Lenz, \jhep {\bf 0605}, 004 (2006),
P.~Ball, V.M.~Braun, Y. Koike, and K.~Tanaka, \npb {\bf 529}, 323 (1998),
P.~Ball, \jhep 9809 (1998) 005.
%Higher-twist distribution amplitudes of the K meson in QCD%
%Higher twist distribution amplitudes of vector mesons in QCD: Formalism and twist - three distributions%
%B ---> pi and B ---> K transitions from QCD sum rules on the light cone%

\bibitem{prd84-034018}
Y.U.~Chun and H.N.~Li, \prd {\bf 84}, 034018 (2011).

\bibitem{plb84-193}
W.~Siegel, \plb {\bf 84}, 193 (1979).
%Supersymmetric Dimensional Regularization via Dimensional Reduction %

\bibitem{epjc40-395}
M.~Nagashima and H.N.~Li, \epjc {\bf 40}, 395 (2005).
%Two parton twist three factorization in perturbative QCD%

\bibitem{jhep0601-067}
J.P.~Ma and Q.~Wang, \jhep {\bf 0601}, 067 (2006); ~~\plb {\bf 642}, 232 (2006).
%Transverse momentum dependent factorization for radiative leptonic decay of B-meson%
%On Transverse-Momentum Dependent Light-Cone Wave Functions of Light Mesons%

\bibitem{plb543-66}
X.Ji and F.Yuan, \plb {\bf 543}, 66 (2002).
%Parton distributions in light cone gauge: Where are the final state interactions?%

\bibitem{appb34-3103}
J.C.~Collins, \appb {\bf 34}, 3103 (2003).
%What exactly is a parton density?%

\bibitem{arxiv1308-0413}
H.N.~Li, arXiv:1308.0413.
%Resummation with Wislson lines off the light cone%

\bibitem{jhep1302-008}
H.N.~Li, Y.L.~Shen and Y.M.~Wang, \jhep {\bf 1302}, 008 (2013).
%Resummation of the rapidity logarithms in B meson wave functions%

\bibitem{arxiv1310-3672}
H.N.~Li, Y.L.~Shen and Y.M.~Wang, arXiv:1310.3672.
%Joint resummation for pion wave function and pion transition form factor%

\bibitem{prd67-034001}
M.~Nagashima and H.N.~Li, \prd {\bf 67}, 034001 (2003).
%k(T) factorization of exclusive processes%


\bibitem{prd71-014015}
P.~Ball and R.~Zwicky, \prd {\bf 71}, 014015 (2005).
%New results on B ---> pi, K, eta decay formfactors from light-cone sum rules%

\bibitem{li2005}
H.N. Li, S. Mishima, and A. I. Sanda, \prd {\bf 72}, 114005 (2005).
%% Resolution to the $B \to \pi K$ puzzle

\bibitem{xiao2008}
H.S. Wang, X. Liu, Z.J. Xiao, L.B. Guo, and C.D. L\"u, \npb {\bf 738}, 243 (2006);
Z.J. Xiao, Z.Q. Zhang, X. Liu, and L.B. Guo, \prd {\bf 78}, 114001 (2008).

\bibitem{xiao2013}
Z.J. Xiao, W.F. Wang, and Y.Y. Fan,  \prd {\bf 85}, 094003 (2012);
W.F. Wang and Z.J. Xiao, \prd {\bf 86}, 114025 (2012);
Y.Y. Fan, W.F. Wang, S. Cheng, and Z.J. Xiao, \prd {\bf 87}, 094003 (2013).

\bibitem{prd67-094013}
Z.T,~Wei and M,Z,~Yang, \prd {\bf 67}, 094013 (2003).
%Phenomenological study of Sudakov effects in the pion form-factor%

\bibitem{prd57-443}
H.N.~Li and B. Tseng, \prd {\bf 57}, 443 (1998).
%5 Nonfactorizable soft gluons in nonleptonic heavy meson decays

\bibitem{prd17-1693}
C.J.Bebek et al, \prd {\bf 17}, 1693 (1978).
%Electroproduction of single pions at low epsilon and a measurement of the pion form-factor up to q^2 = 10-GeV^2%

\bibitem{prc78-045203}
G.M.Huber et al, (Jefferson Lab Collaboration), \prc {\bf 78}, 045203 (2008).
%Charged pion form-factor between Q**2 = 0.60-GeV**2 and 2.45-GeV**2. II. Determination of, and results for, the pion form-factor%


\end{thebibliography}
\end{document}